\documentclass[aps,prl,reprint,groupedaddress,amsmath,amssymb,byrevtex]{revtex4-2}
\usepackage{graphicx}
\usepackage{color}
\definecolor{colortodo}{RGB}{255,0,0}

\definecolor{colortodo2}{RGB}{0,255,0}

\begin{document}

\title{Streamwise dissolution patterns created by a flowing water film}

\author{Adrien Gu\'erin}
\affiliation{MSC, Univ Paris Diderot, CNRS (UMR 7057), 75013 Paris, France}
\author{Julien Derr}
\affiliation{MSC, Univ Paris Diderot, CNRS (UMR 7057), 75013 Paris, France}
\author{Sylvain Courrech du Pont}
\affiliation{MSC, Univ Paris Diderot, CNRS (UMR 7057), 75013 Paris, France}
\author{Michael Berhanu } 
\email{michael.berhanu@univ-paris-diderot.fr}
\affiliation{MSC, Univ Paris Diderot, CNRS (UMR 7057), 75013 Paris, France}

\date{\today}

\begin{abstract}
The dissolution of rocks by rainfall commonly generates streamwise parallel channels, yet the occurrence of these natural patterns remains to be understood. Here, we report the emergence, in the laboratory, of a streamwise dissolution pattern at the surface of an initially flat soluble material, inclined and subjected to a thin runoff water flow. Nearly parallel grooves of width about one millimeter and directed along the main slope spontaneously form. Their width and depth increase continuously with time, until their crests emerge and channelize the flow. Our observations may constitute the early stage of the patterns observed in the field.

\end{abstract}

\maketitle

\begin{figure}
  \includegraphics[width=.9\columnwidth]{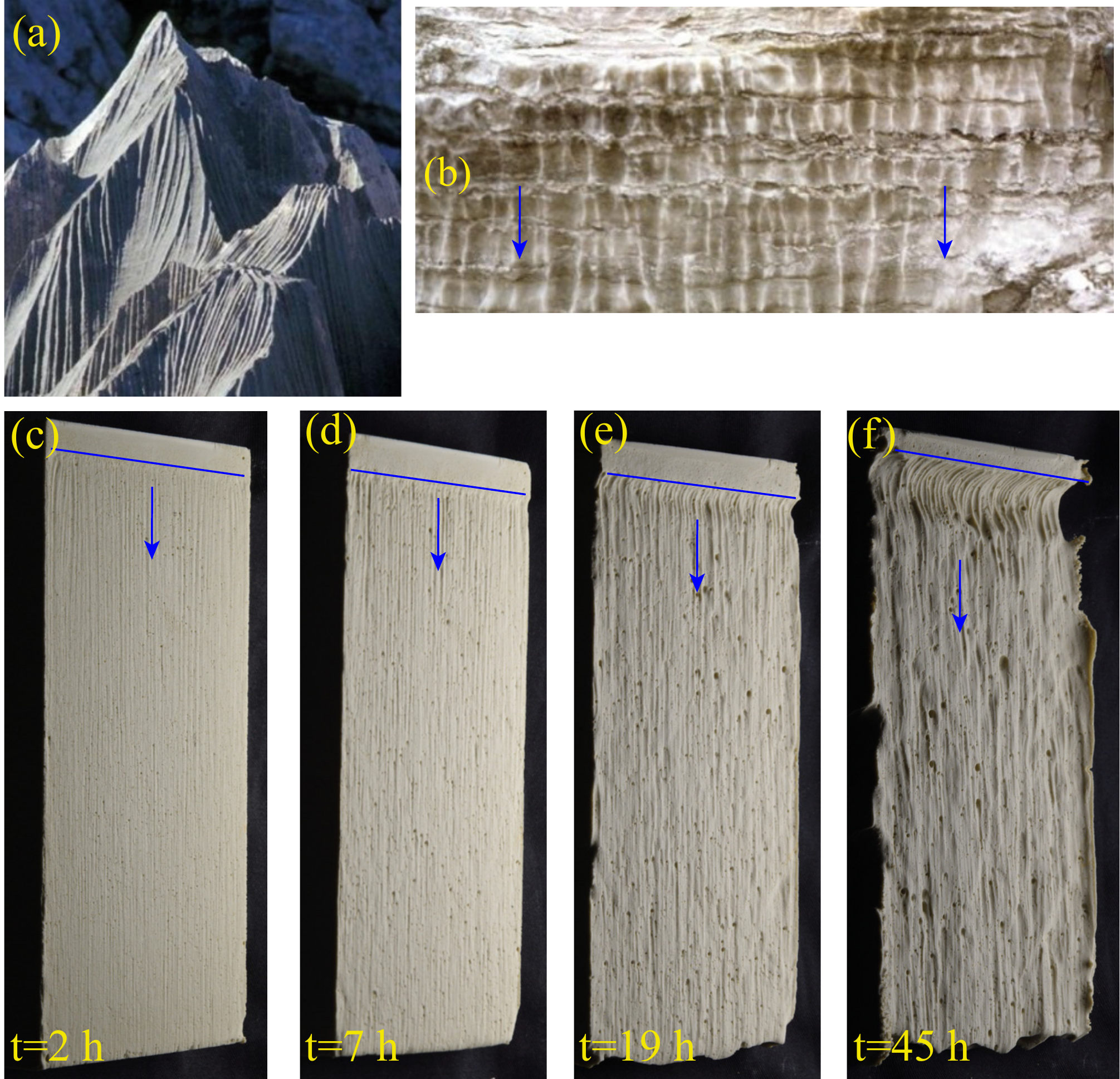}
\caption{(a) Natural \textit{Rillenkarren} on limestone (Karst plateau, Slovenia),  Credit Marko Simic. Width of the picture about 1 m.  (b) Dissolution rills or \textit{Rillenkarren} on Gypsum (Vaucluse, France) Width of the picture about 20 cm. Credit Pierre Thomas (https://planet-terre.ens-lyon.fr). (c-d-e-f) Experimental dissolution grooves. Consecutive pictures of a gypsum (plaster of Paris) block subjected to a runoff flow (inclination $39$ deg, Flowrate $Q=2.8$ L/min, $U=0.84$ m/s) for increasing times of exposure to the flow. From left to right, $t=2$, 7, 19 and 45 hours. The blue lines indicate the position of water injection and the blue arrows the flow direction {(b-f)}. The initial dimensions of the block are $200 \times 100 \times 25$\,mm.}
 \label{Fig1}
\end{figure}

Pattern formation arising from mechanical erosion has been the subject of numerous physics studies~\cite{Charru2013,Courrech2015,Jerolmack2019}. In contrast, morphogenesis due to chemical erosion has received far less attention outside the geological community~\cite{Meakin2009,Jamtveit2012}. Yet, it is the dominant erosive process for soluble rocks such as limestone ($CaCO_3$), gypsum ($CaSO_4,\,2 H_2O$) and salt ($NaCl$). When the rock is in contact with a water flow, it dissolves into the flow and becomes a solute. The rock erosion rate then depends on the local solute concentration at the dissolving interface. {This concentration can vary due to hydrodynamics, explaining the emergence of specific dissolution patterns~\cite{Meakin2009,Jamtveit2012}}. In caves, solutal convection{~\cite{Sullivan96,Cohen2016,Wykes2018, Cohen2020,Pegler2020}} and turbulent flows~{\cite{Huang2015,Moore2017,Blumberg1974,Claudin2017}} shape dissolution pits and scallops. At the Earth surface, when the rock is exposed to rainfall, the runoff flow typically creates nearly parallel channels, called \textit{Rillenkarren} in the geomorphology literature. They are directed along the main slope and their width ranges from about one millimeter to one meter~\cite{Karrenbook,LUNDBERG2013121} (see Fig.~\ref{Fig1} (a-b)). Although their occurrence is common in nature, the physical origin of this streamwise pattern remains unknown.

At the surface of a granular bed, streamwise grooves can be generated by counter-rotating vortices when the flow is turbulent{~\cite{Colombini1993,Colombini1995}}, or by the diffusion-like transport of grains from the troughs to the crests of the bed in laminar regime~\cite{Abramian2019}. Yet, when water flows over a melting, dissolving or precipitating bed, stability analyses only predict spanwise (or transverse) undulations, due to a coupling between the free surface of the flow and the bed~\cite{Camporeale2012,Vesipa2015,Yokokawa2016}. Streamwise patterns on soluble rocks have only been predicted for thin film flows overhung by the rock, which are destabilized by the gravity~\cite{Camporeale2015,Bertagni2017,Balestra2018}. 

Experimentally, turbulent flows of about $0.5$ m/s in a deep-water regime generate streamwise grooves called flutes on plates of plaster~\cite{Allen1970} (same chemical composition as gypsum). This streamwise dissolution pattern can evolve into a transverse pattern commonly called scallops, {but the author does not monitor the pattern change and proposes only qualitative mechanisms.} Finally, a single study reports the formation of streamwise grooves in experimental conditions close to the \textit{Rillenkarren} ones~\cite{Glew1977,Glew1980}. Molded blocks of plaster and salt blocks were subjected to an artificial rain generating a flow of a few cm/s and around $100\,\mu$m deep. After several hundred hours, several-centimeters-wide grooves were obtained. This descriptive study does not investigate the initiation nor the temporal evolution of this pattern. Several authors claim that the impact of the rain droplets on the water film is the condition to generate the grooves~\cite{Glew1980,Lundberg2009,LUNDBERG2013121}.

Here, we demonstrate experimentally the emergence of streamwise dissolution patterns created by a water film flowing on soluble rocks, in the absence of rainfall. Fig.~\ref{Fig1} (c) to (f) show the results of an experiment performed on a block of plaster of Paris (KRONE Alabaster Modellgips 80, pure gypsum, see Sup. Mat.~\cite{Supp} which includes 
Refs.~\cite{Fallingfilmsbook,CharruBook,PIVlab,Liu1993,Colombani2007,Handbook,Alkattan,crank1979mathematics}). Millimeter-wide grooves directed along the main slope appear after several dozens of minutes. Over time, the characteristic width and depth of the grooves increase, until they are so deep that their crests emerge from the water film, thus channelizing the flow. At this stage, these centimetric grooves are similar to the \textit{Rillenkarren} observed in the field.

Experiments with limestone are difficult to perform in reasonable times, as limestone dissolves slowly. Yet, we also performed experiments on carved blocks of Himalaya pink salt (Khewra Salt Mine, Pakistan, 98\% of sodium chloride~\cite{Sharif2007}). Although these blocks are natural geological samples and present a significant heterogeneity in structure, with chemical defects and cracks, we observed qualitatively similar streamwise grooves (Sup. Mat.~\cite{Supp}). 

\begin{figure}

   \includegraphics[width=\columnwidth]{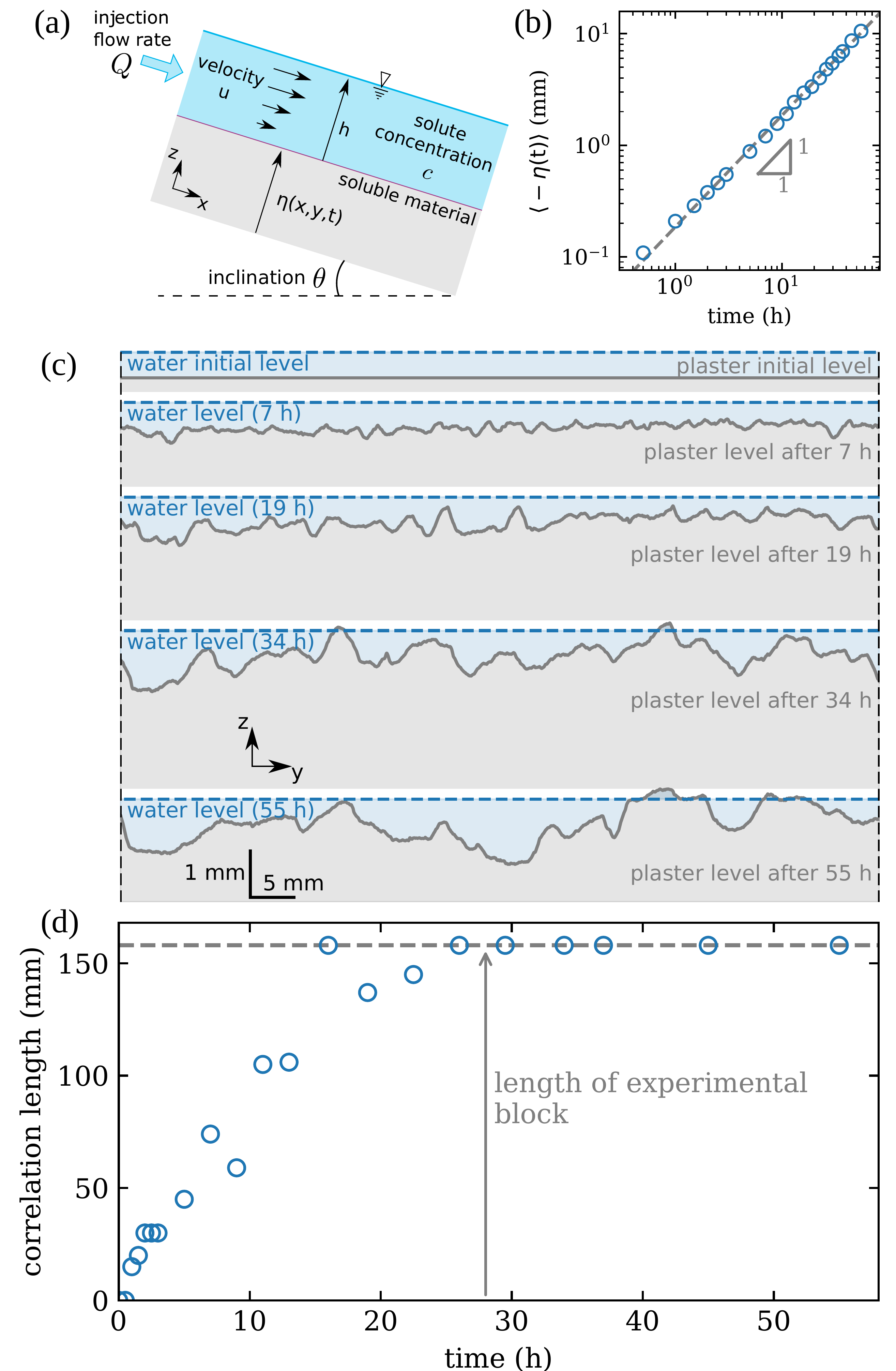}

 \caption{ (a) Sketch of the dissolution experiment, side view. A water film of flow rate $Q$, velocity $U$ and thickness $h$ is driven by gravity and flows over a block of soluble rock of width $W=100\,$mm, length $L=200\,$mm, and inclined at an angle $\theta$. The block thickness $\eta(x,y,t)$ progressively decreases as it dissolves, creating a solute concentration field $c(x,y,z)$ in the flow. (b) As a result, the surface-averaged dissolved depth $\langle -\eta \rangle$ increases linearly (here during a typical experiment with $U=0.84\,$m/s, $h=380\,\mu$m). (c) Transverse profiles $\eta(x,y,t)$ of the same experiment, measured along $y$ at a distance $x=20\,$mm from the water injection, and plotted at various times $t$. The water level is deduced by flow conservation. Thin grooves grow over an initially flat surface, and finally form channels, their crests emerging from the flow. (d) As the streamwise pattern grows, the longitudinal correlation length, which measures the maximum distance along $x$ between two correlated transverse profiles, increases, until it reaches the length of the block.}
 \label{Fig2}
 \end{figure}

Fig.~\ref{Fig2}~(a) depicts the principle of the experiment. A constant flowrate $Q$ of tap water is injected on top of molded rectangular blocks of plaster ($200\times 100\,\times 25\,$mm). Thanks to the excellent wetting properties of soluble materials, a thin water film of homogeneous thickness $h$ spans all the block after a transient of a few seconds. Driven by gravity, the film flows over the top surface of the block inclined with an angle $\theta$. We measure the average film thickness $h$ with the flight time of an ultrasonic beam, and then deduce the average velocity $U$ (Sup. Mat.~\cite{Supp}), {for the initial flat bed}. For example, with a typical flowrate $Q=2.8$ L/min and an inclination $\theta =39$ deg, we measure a thickness $h=560\,\mu$m, and calculate a velocity $U \approx 0.84\,$m/s. The corresponding Reynolds and Froude number are $Re=(U\,h)/\nu \approx 466$ and $Fr=U/\sqrt{g\,h} \approx 11$ (with $g=9.81$m s$^{-2}$ the gravitational acceleration and $\nu$ the kinematic viscosity of fresh water  $\nu=1.0\,10^{-6}\, $ m$^2$ s$^{-1}$ at $20^\circ$ C). According to experimental~\cite{Ishigai1972} and numerical works~\cite{Dietze2014}, these thin flowing films belong to a regime of wall-induced turbulence ($Re > 75$).

Measuring the topography of a rock while a thin film of water flows over it is a difficult task. Instead, we regularly stop the experiment and measure the three-dimensional topography of the dried eroded surface using a laser scanner. We thus obtain the surface elevation $\eta (x,y,t)$ with an accuracy of $70\,\mu$m and a resolution in $(x,y)$ of $0.2$\,mm at several time steps, ranging from 0.5 to 10 hours. We can then analyse the topography evolution of the plaster blocks. As an example, let us describe a typical, 55-hours-long experiment with $U=0.84$\,m/s and $h=560$\,$\mu$m. As water dissolves the gypsum and flows downwards, the solute concentration increases. The erosion resulting from this dissolution is thus slower far from the injection, as the solute concentration gets closer to its saturation value. The erosion dynamics and the resulting shaping of longitudinal profiles will be studied in a further work, and we only stress here that the surface-averaged erosion $\langle -\eta(t) \rangle$ of the rock is linear in time, which means that the erosion rate of the block is constant (Fig.~\ref{Fig2}~(b)). 

After 30 minutes of exposure to the flow, streamwise thin grooves become visible. In Fig.~\ref{Fig2} (c), consecutive transverse profiles $\eta(x,y,t)$ represent the progressive erosion of the rock at a distance $x=20\,$mm from the water injection. The associated water height deduced by flow conservation is also displayed. As time grows, both the grooves' amplitude and width increase. The thin grooves progressively merge into the large ones, which eventually channelize the flow.

At the beginning, the grooves are a few centimeters long and randomly distributed over the rock surface. Then, they progressively extend all along the block. We measure the streamwise coherence of the pattern by measuring the cross-correlation coefficient between transverse profiles. As the distance $d$ separating the profiles along the $x$-axis grows, the correlation coefficient decreases. We define the correlation length as the distance $d$ where the correlation coefficient becomes smaller than a threshold, chosen to be 0.2 (Supp. Mat.~\cite{Supp}). Fig.~\ref{Fig2} (d) shows that the correlation length increases with time, and eventually reaches, after about 25 hours, a saturation value corresponding to the sample size. Then, the grooves span the whole length of the block.

 \begin{figure}
 \begin{center}

 \includegraphics[width=1\columnwidth]{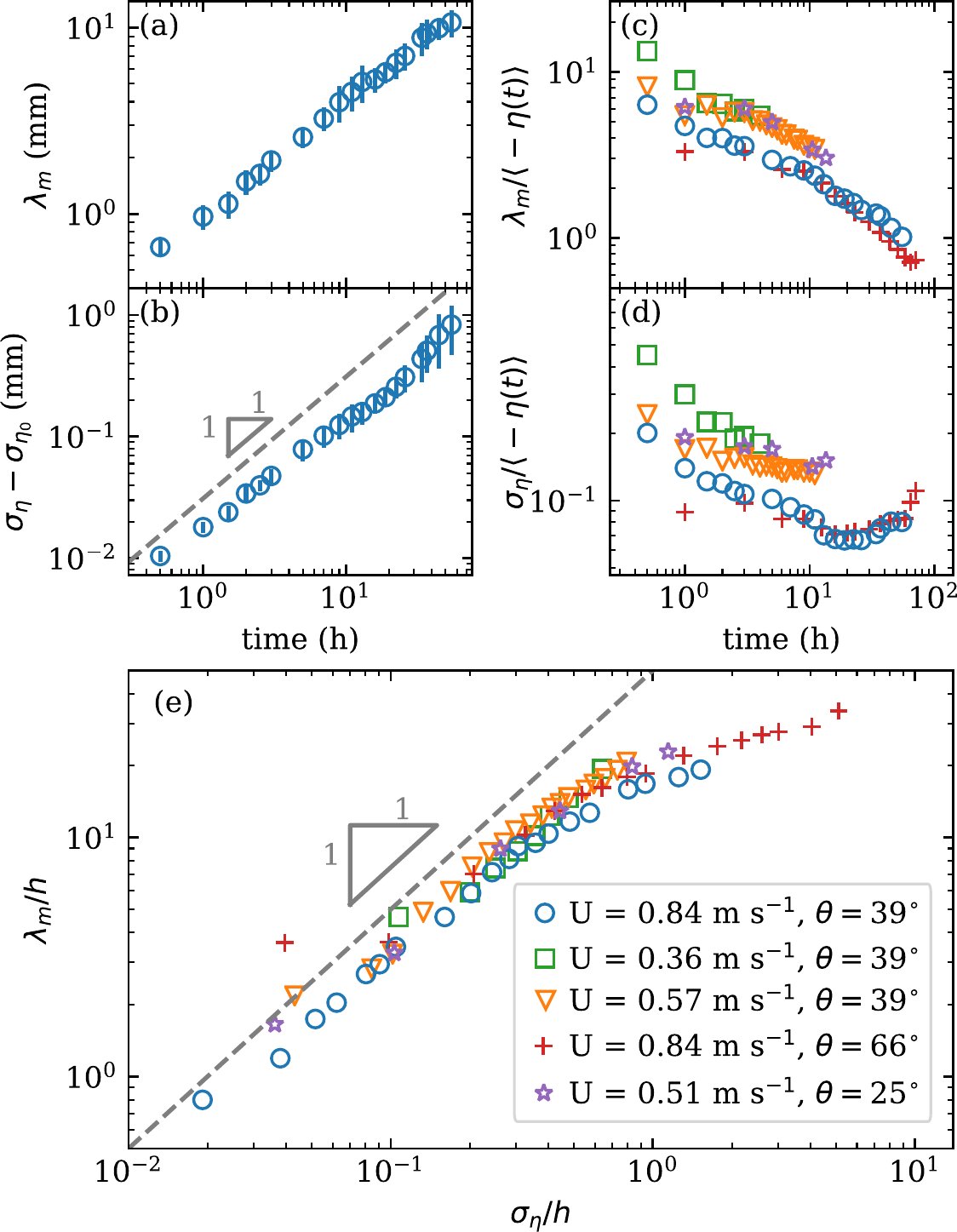}
  \end{center}
 \caption{Evolution of grooves morphology during five experiments.  Blue $\circ$, $Q=2.8$ L/min, $\theta=39$ deg, $U=0.84$ m/s and $h=556$ $\mu$m. Green $\square$, $Q=0.22$ L/min, $\theta=39$ deg, $U=0.36$ m/s and $h=101$ $\mu$m. Orange $\triangledown$, $Q=0.86$ L/min, $\theta=39$ deg, $U=0.57$ m/s and $h=255$ $\mu$m. Red $+$, $Q=1.9$ L/min, $\theta=66$ deg, $U=0.84$ m/s and $h=380$ $\mu$m. Magenta $\star$, $Q=0.93$ L/min, $\theta=25$ deg, $U=0.51$ m/s and $h=305$ $\mu$m. (a) The groove typical wavelength $\lambda_m$ increases with time. (b) The pattern amplitude $\sigma_\eta (t)-\sigma_\eta (t=0) $ grows linearly with time. {The error-bars display the standard deviation along the $x$ axis.}  (c) Grooves typical wavelength rescaled by the dissolved depth $\lambda_m/|\langle  \eta \rangle |$ versus time. (d) Pattern amplitude rescaled by the dissolved depth $\sigma_\eta/|\langle  \eta \rangle |$ versus time. (e) Grooves wavelength with respect to their amplitude, both rescaled by the water depth. The aspect ratio seems to be controlled by the flow depth.}
  \label{Fig3}
  \end{figure}

We measure the typical wavelength of a profile by computing the weighted average of its Fourier spectrum (Sup. Mat.~\cite{Supp}). As it is constant over the length of the block, we estimate the grooves' typical width $\lambda_m$ as the average profiles wavelength. Fig.~\ref{Fig3}~(a) shows that it grows during the whole experiment, from about $0.6$\,mm to $10$\,mm in 55 hours. The pattern amplitude, estimated by the profile standard deviation $\sigma_\eta$ minus the initial rugosity, and averaged over the length of the block, also increases during the whole experiment, and almost linearly (Fig.~\ref{Fig3}~(b)).

We now analyze five experiments performed on plaster, with velocities varying from $0.36$ to $0.84\,$m/s and three block inclinations of 25, $39$ and $66$ deg (Supp. Mat.~\cite{Supp}). 
Fig.~\ref{Fig3}~(c) (and (d) respectively) show the grooves width (resp. their amplitude) divided by the eroded depth $\langle -\eta(t)  \rangle$. With this rescaling, the data { belonging to different runs are gathered.}

Finally, the grooves aspect ratio seems to be controlled by the water depth. Indeed, Fig.~\ref{Fig3} (e) represents $\lambda_m$ as a function of $\sigma_\eta$, both rescaled by the water depth $h$, which is the natural lengthscale of the problem. The different points obtained at different times and several velocities all collapse on to the same curve. The initial value of $\lambda_m/h$ varies from $0.8$ to $1.5$, showing that the first pattern wavelength is of order of the water depth. Then, as long as $\sigma_\eta/h \lesssim 1$, $\lambda_m$ and $\sigma_\eta$ evolve proportionally, with $\lambda_m \approx 20\, \sigma_\eta$. When the pattern amplitude exceeds the water height, \textit{i.e.} $\sigma_\eta \gtrsim h $, $\lambda$ evolves more slowly than $\sigma_\eta$, meaning that the grooves aspect ratio is no longer conserved. In this channelization regime, the channel crests start to emerge from the water film. The flow thus keeps on dissolving the troughs while the crests are more preserved, such that the channels depth increases faster than their width.\\

 \begin{figure}
 \begin{center}
 \includegraphics[width=.75\columnwidth]{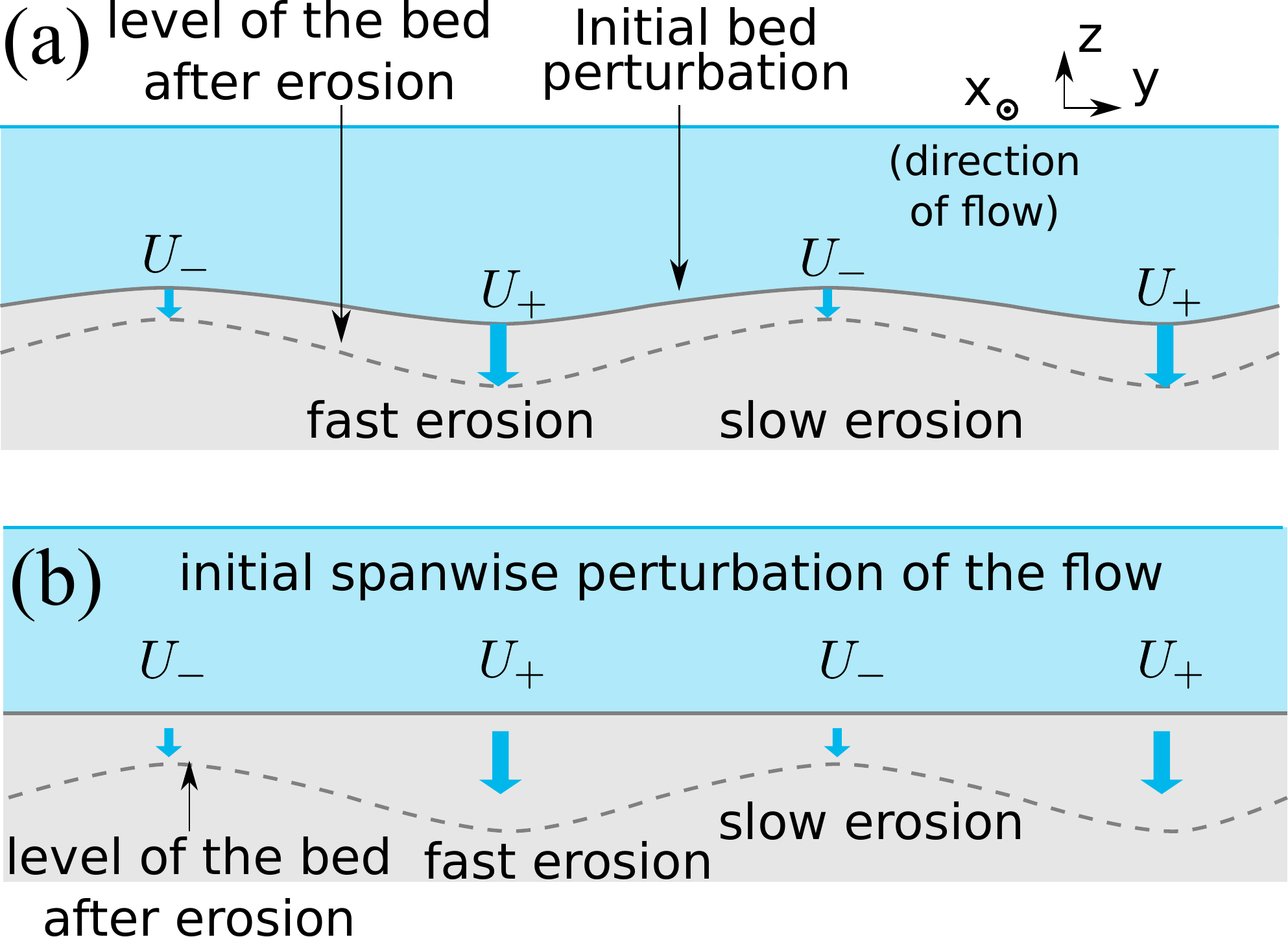}
  \end{center}
 \caption{Two scenarios of apparition of a streamwise dissolution pattern. (a) An initial perturbation of the bed is amplified, because the local velocity increases with the fluid depth for a gravity driven flow. (b) The flow contains a transverse velocity variation, which is imprinted on the bed.}
 \label{Fig4}
 \end{figure} 
 
\paragraph*{Discussion.} In this experimental study, we demonstrate that streamwise dissolution patterns are created by a water runoff flow of constant thickness. This observation contradicts the affirmation that the impact of rain drops is required to create the dissolution rills observed in the field~\cite{Glew1980}. To our knowledge, such streamwise dissolution patterns have not yet been predicted theoretically.

Dissolution patterns result from a spatial heterogeneity of the erosion rate, which can be caused by local variations of the flow velocity. Indeed, a larger velocity decreases the thickness of the solute concentration boundary layer at the dissolving interface, which increases the local erosion rate. Thus we propose two kinds of mechanism to explain the pattern emergence.

First, grooves could stem from a feedback mechanism between the topography and the flow, leading to a destabilization of the bed. For a gravity-driven flow, the local velocity increases with the water depth. Thus, any initial perturbation of the bed would be amplified, as displayed in Fig.~\ref{Fig4} (a). A simple linear stability analysis in the spanwise plane indeed predicts that all wavelengths are unstable {and an exponential growth} (Sup. Mat.~\cite{Supp}). {However, experimentally the pattern grows linearly at short times. Moreover,} the initial grooves width could be imposed by the defects of the dissolving surface like small-scale roughness or chemical heterogeneities. Yet, the grooves morphology is similar on salt and plaster experiments (see pictures in Sup. Mat.~\cite{Supp}), and on gypsum and limestone (Fig.~\ref{Fig1} (a,b)). The regularity of their morphology on various materials suggests that this pattern does not depend on the material heterogeneities, which vary from one material to the other, but rather on a purely hydrodynamic mechanism.

Patterns could also arise because of the hydrodynamics only, being an imprint of heterogeneities in the flow. If the velocity profile presents transverse variations, as shown in Fig.~\ref{Fig4} (b), the pattern would then appear as a passive response of the dissolving bed to the flow, which would impose the initial wavelength $\lambda_m$. A similar scenario has been proposed for the emergence of dissolution patterns in solutal convection~\cite{Philippi2019,Cohen2020}. Here, by taking for example $U=U_0\,[1+\beta \, \cos(k\,y)]$ and modeling simply the solute advection, we predict that the pattern evolution velocity $\mathrm{d}\sigma_\eta/\mathrm{d}t$ is proportional to the global erosion velocity $\mathrm{d} |\langle \eta(t)  \rangle| /\mathrm{d}t$ by a factor $\beta/2$ (Sup. Mat.~\cite{Supp}). Although justified only at short times, this model implies a linear growth of the pattern amplitude and a constant value of the ratio $\sigma_\eta/|\langle \eta \rangle |$, which are both observed in our experiments (Fig.\ref{Fig3}~(b,d)). However, we do not know of such hydrodynamic structures reported in the literature, such that $\beta$ and $k$ are unknown. They must have a sufficient lifetime to have a significant influence, as the erosion velocity is several orders of magnitude smaller than the flow velocity. For a turbulent, deep water flow, Allen explains qualitatively the apparition of grooves and furrows on plaster blocks by the presence of turbulent coherent structures~\cite{Allen1970}, like the turbulent streaks~\cite{Kline1967,CHERNYSHENKO2005} inside the viscous sublayer or the G\"ortler vortices~\cite{Saric1994} due to the substrate curvature. After initiation, the structure locations could be locked at the positions of the first ridges, which could fix the phase of the pattern during its evolution. A decisive proof of the existence of these structures in the conditions of our experiment is difficult to obtain, as hydrodynamic measurements in films thiner than one millimeter are challenging. In the conditions of our experiments, the flow belongs to a wall-induced turbulence regime~\cite{Ishigai1972,Dietze2014}, and the presence of turbulent structures cannot be excluded. Yet, the wavelength of these structure are predicted theoretically only in semi-infinite domains. Moreover, such structures have never been reported in shallow flows, and we must note the absence of any velocity field characterization of turbulent falling films, neither experimentally nor numerically~\cite{Dietze2014}. Our experimental results suggest that such structures may exist, and thus stress that the hydrodynamics of falling films at large Reynolds numbers still present unknown features which could strongly affect mass and heat fluxes. The film thickness $h$ is a natural lengthscale and seems to control the morphology of the grooves observed in our experiment. The emergence and shape of flow structures into a thin flowing film probably requires to take full account of this lengthscale.

In the field, the \textit{Rillenkarren} are created by a rainfall-generated flow, which presents significantly smaller depth and velocity than our experiments~\cite{Glew1980}. However, the impacts of millimetric drops could efficiently induce small-scale turbulence~\cite{Harrison2017} and generate spanwise perturbations to the velocity field. Once they have emerged at the earliest stages of the experiment, the grooves then grow continuously in length, width and depth. This suggests that the \textit{Rillenkarren} observed in the field are probably a snapshot of a continuous evolution. If this is the case, the \textit{Rillenkarren} dimensions would indicate their age, that is their time of exposure to rainfall.

Finally, the long-term evolution of the pattern deserves further study. Our results show that, once the grooves' depth becomes comparable to the flow depth, their crests emerge from the flow and the grooves channelize the flow. Then, they enter a new growth regime where their width grows much more slowly than their depth. Understanding this growth regime would need to take into account the feedback of the evolving topography on the flow. In deep-water flows, the dissolution patterns tend to evolve into traverse patterns called scallops~\cite{Allen1970,Cohen2020}. Here, the flow channelization may prevent the development of such transverse instabilities.
\\

\begin{acknowledgments}
We thank Valentin Leroy for his help with the ultrasonic measurements, Alexandre Di Palma for technical assistance {and Jean-Gabriel Pichon for experimental work with salt blocks}. We acknowledge Piotr Szymczak, Matteo Bertagni and Carlo Camporeale for discussions. This research was funded by the ANR grant Erodiss ANR-16-CE30-0005.
\end{acknowledgments}


%

\end{document}


\title{Supplemental Material.\\ ``Streamwise dissolution patterns created by a flowing water film."}

\author{Adrien Gu\'erin }
\affiliation{MSC, Univ Paris Diderot, CNRS (UMR 7057), 75013 Paris, France}
\author{Julien Derr}
\affiliation{MSC, Univ Paris Diderot, CNRS (UMR 7057), 75013 Paris, France}
\author{Sylvain Courrech du Pont}
\affiliation{MSC, Univ Paris Diderot, CNRS (UMR 7057), 75013 Paris, France}
\author{Michael Berhanu } 
\email{michael.berhanu@univ-paris-diderot.fr}
\affiliation{MSC, Univ Paris Diderot, CNRS (UMR 7057), 75013 Paris, France}

\date{\today}

\maketitle

%
%

In this supplemental material, we provide additional information about the experimental setup (\S~\ref{expsetup}) and the flow characterization (\S~\ref{flowcha}). The methods for surface analysis are presented in \S~\ref{surfchar}. We show also characteristic pictures of patterned plaster blocks and a table summarizing the plaster experiments (\S~\ref{expruns}). Finally, we present in \S~\ref{linstab} a simplified linear stability analysis, which is compared to the passive response of the bed to transverse flow perturbations in  \S~\ref{response}. Notations as in the aforementioned paper.


\section{Experimental setup}
\label{expsetup}

\begin{figure}
 \begin{center}
   \includegraphics[width=.4\linewidth]{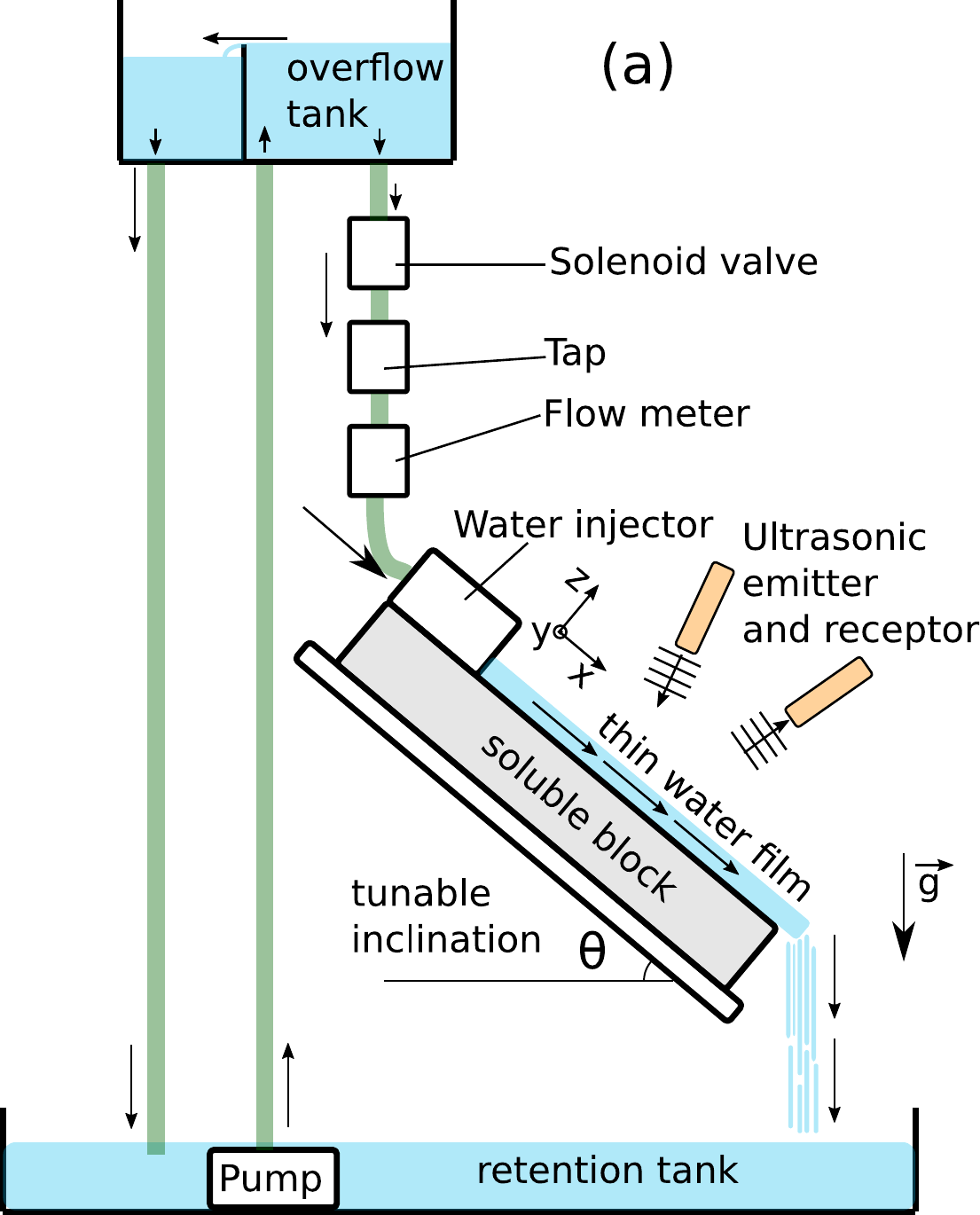} \hfill 
  \includegraphics[width=.58\linewidth]{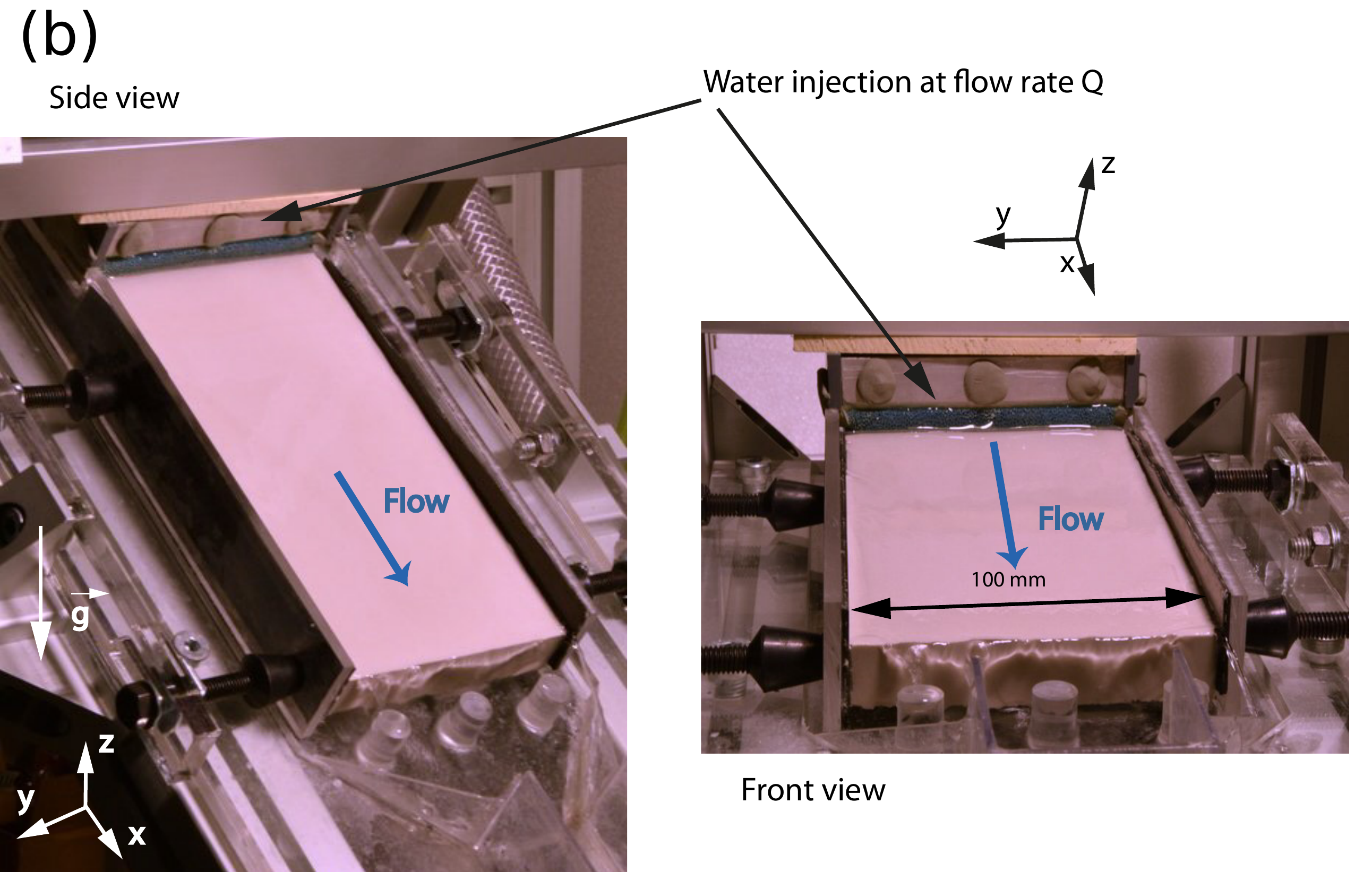} 
 \end{center}
 \caption{(a)  Sketch of the experimental setup, side view, not to scale. (b) Pictures of the experimental setup used with a plaster block.}
 \label{FigSetup}
 \end{figure}

We performed experiments on two types of soluble rocks, namely plaster of Paris and salt. The plaster, made of pure gypsum, is a 2/3 - 1/3 mixture of plaster powder (KRONE Alabaster Modellgips 80) and tap water, poured into an acrylic mold of dimensions $200\times 100\,\times 25\,$mm. The bottom of the mold imprints its surface state on the plaster block, which surface is consequently perfectly flat and smooth. The salt blocks are extracted from the Khewra Salt Mine, in Pakistan, and sold at approximately the same dimensions ($200\times 100\,\times 25\,$mm$^3$). Before subjecting a block to a water flow, we sand its top surface with an emery cloth, such that the surface be flat and approximately smooth (it adopts the roughness of the emery cloth). We carried out also one unique experiment with a polished plate of Alabaster provided complimentary by the \textit{Atelier Alain Ellouz} ( https://atelier-alain-ellouz.com/ ). Alabaster is a compact rock made of gypsum and has the same chemical composition as the plaster blocks.

The experimental setup is depicted in Fig.~\ref{FigSetup}. The block is inclined at an adjustable angle between $0$ and $90$~deg, and maintained laterally by rubber sheets for sealing. A custom-made water injector is placed directly over the top surface of the rock, over the first 20 millimeters of its upper left part, thus preserving the first 20 millimeters of this surface from the flow. The rock surface actually subjected to the flow is consequently about $180\times 100\,$mm. In order to insure a constant flow rate, we use an overflow tank located about one meter above the sample. This tank is connected to the water injector, which outlet is buffered with an aquarium foam to ensure that the water injection is homogeneous over the whole width of the sample. Thanks to the excellent wetting of soluble materials, the water exiting this injector then flows over the rock and forms a film of homogeneous thickness over the whole rock surface after a transient of less than ten seconds. The water then escapes to a retention tank {of 120 L}. Salt experiments were only 1 to 4 minutes long, and we only used fresh tap water. Plaster experiments were 4 to 71 hours long, and we could not afford to use only fresh water. The already-used water in the retention tank was therefore sent back to the overflow tank, where it was re-used for the experiment, such that the water circulated in a closed loop. In order to avoid an increase of solute concentration in the flow and possible saturation, the whole water volume was then completely renewed periodically (every 1 to 4 hours), such that its solute concentration was always below 0.1-0.2\,g\,l$^{-1}$, that is ten to twenty times lower than the saturation concentration. 


Finally, the flowrate $Q$ through the injector is controlled using a tap, and monitored with a digital flowmeter. The duration of exposure to the water flow is precisely set using a solenoid valve interfaced with Arduino.

\section{Flow characterization}
\label{flowcha}

The fluid mechanics of falling films constitutes a vibrant field of research due to the experimental challenge to characterize the velocity field in a film of thickness smaller than one millimeter, and to the intrinsic nonlinear dynamics of the free-surface~\cite{Fallingfilmsbook}. In our experiment, the control parameters are the imposed flowrate $Q$ and the block inclination $\theta$. In stationary regime and with the hypothesis of homogeneous flow, these parameters determine the film thickness $h$ and the average velocity $U$ as a balance between wall friction and the hydrostatic pressure gradient caused by the gravity $\mathbf{g}$. The laminar solution in viscous regime, the Nusselt solution, corresponds to a half-parabolic profile. Then, $Q$, $h$ and $U$ read~\cite{CharruBook,Fallingfilmsbook} for a film of width $W$ with $\nu$ the kinematic viscosity:
\begin{eqnarray}
Q&=&U\,h\,W \\
h&=&\left(\dfrac{3\nu\,Q}{W\,g\,\sin \theta} \right)^{1/3}  \label{hNu}\\
U&=&\left(\dfrac{Q^2\,g\,\sin \theta}{3\,\nu\,W^2} \right)^{1/3}  \label{UNu}
\end{eqnarray}\\

  \begin{figure}[h!]
 \begin{center}
  \includegraphics[width=.4\linewidth]{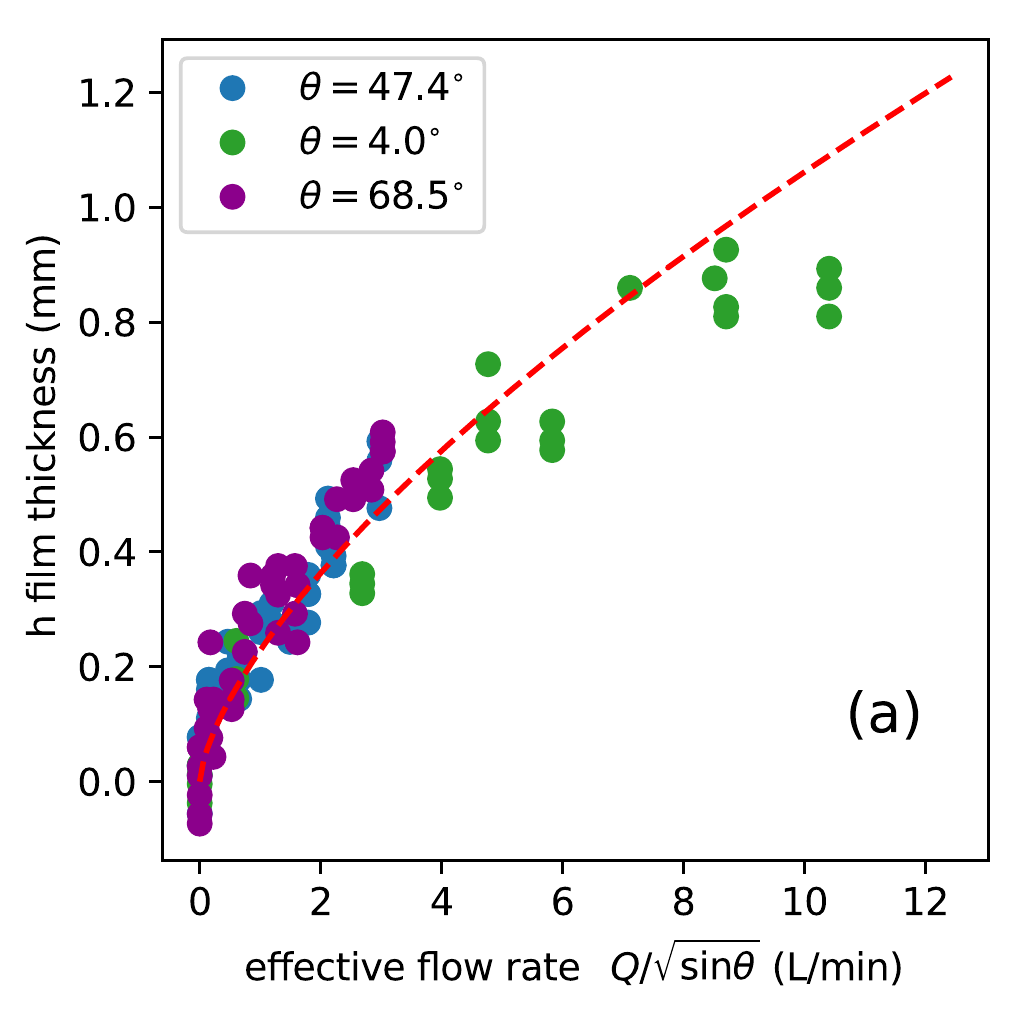} \hfill
  \includegraphics[width=.52\linewidth]{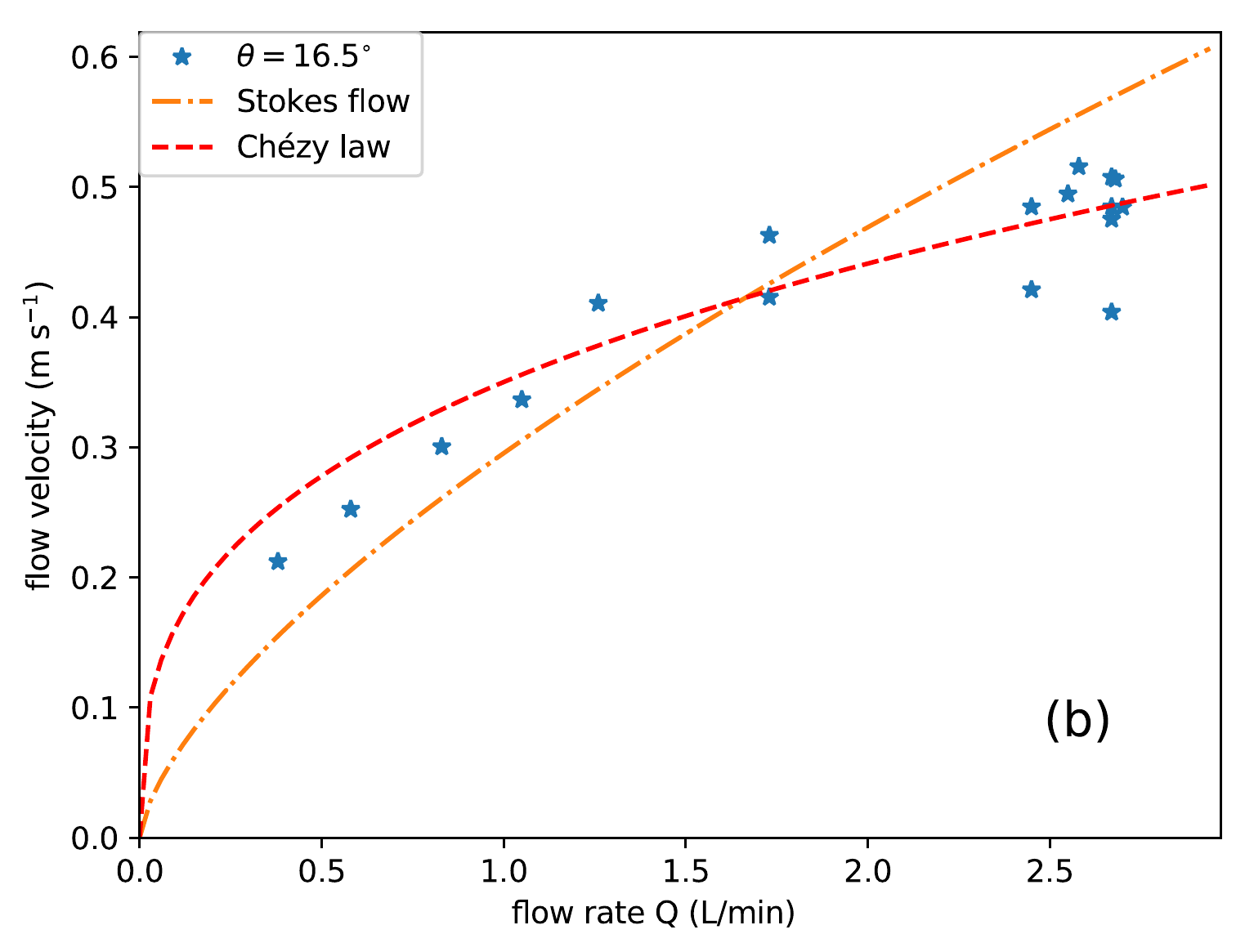}
 \end{center}
 \caption{ (a) Thickness of the water film as a function of the effective flow rate $Q/\sqrt{\sin \theta}$ for three different inclinations $\theta$, and measured by the flight time of an ultrasonic beam. Dashed red line, prediction for a gravity-driven  flow balanced by a friction term given by the Ch\'ezy law of coefficient $C_f=4.5\,10^{-3}$ (Eq.~\ref{hChe}). (b) Time-averaged velocity $U$ in the film, using particle image velocimetry for several values of the flowrate $Q$ and an inclination $\theta=16.5^\circ$. The experimental points are compared to the theoretical estimates given by Eq.~(\ref{UNu}) (orange dashed dot line, viscous solution) and Eq.~(\ref{UChe}) (red dashed line, inertial solution using the Ch\'ezy law).}
 \label{FigFlow}
 \end{figure}

At larger Reynolds numbers, the wall friction results from an inertial drag instead. A simple and empirical approach consists in using equations integrated over the film thickness, the so-called Saint-Venant approximation. {The one dimensional equation for the averaged momentum balance reads~\cite{Fallingfilmsbook} :
$$\partial_t \, (h\,U)+\partial_x \, (h\,U^2)=g\,h\,\sin \theta - \tau_w/\rho - g\,h \cos \theta \partial_x h  \, , $$
where $\partial_t$ means $(\partial \,\,)/(\partial t)$.
We use the Ch\'ezy law modeling the wall drag as $\tau_w/\rho = C_f\,U^2 $, which defines the dimensionless friction coefficient $C_f$.} One obtains in stationary regime for a flat film at the equilibrium:
\begin{eqnarray}
Q&=&U\,h\,W \\
h&=&\left(\dfrac{C_f\,Q^2}{W^2\,g\,\sin \theta} \right)^{1/3}  \label{hChe}\\
U&=&\left(\dfrac{Q\,g\,\sin \theta}{C_f\,W} \right)^{1/3}  \label{UChe}
\end{eqnarray}\\

To characterize the flow in our experiment, the film thickness $h$ has been determined above plaster blocks by measuring the flight time difference of an ultrasonic beam of frequency $40\,$ kHz with and without the flow. {This set of measurements are performed for flat or nearly flat bed and thus correspond to the initial state of the experiment before the development of significant patterns}. The results are displayed in Fig.~\ref{FigFlow} (a) as a function of $Q$ for selected inclinations after time-averaging. $h$ follows fairly a power-law in $Q^{2/3}$ incompatible with the low-Reynolds solution Eq.~(\ref{hNu}), and favoring the inertial solution. $h$ is indeed well described by Eq.~(\ref{hChe}) with $C_f \approx 4.5\,10^{-3}$. Using this relation, we can then deduce the average velocity $U$ in the flow. We were not able to measure the film thickness for all experiments. We therefore use equations~(\ref{hChe}) and (\ref{UChe}) to infer the film thickness and velocity from the flow rate and inclination.

This indirect measurement of the average flow velocity is confirmed using Particle Image Velocimetry (PIV) for a particular set of flowrates for an inclination of $16.5^\circ $ in a domain $x\in [65,90]$ mm from the injector. The liquid film is lighted by a laser sheet using a continuous laser of 2W.  The top surface is imaged from above using a fast camera Phantom V7 with a frame-rate of $1\,$ kHz. Images are processed using PIVlab~\cite{PIVlab}. Due to the thickness of the laser sheet (about $1$ mm), only small inclinations, corresponding to larger film depths, can be tested with this method. Moreover, variations along the $z$ axis, perpendicular to the plaster block, are not resolved. We assumed that the measured velocity corresponds to the velocity $U$ averaged over the thickness. In Fig.~\ref{FigFlow}~(b), we plot the time-averaged velocity  as a function of the flow rate $Q$. Our data is satisfyingly described by the inertial solution with the Ch\'ezy law (Eq.~\ref{UChe}) instead of the Nusselt expression (Eq.~\ref{UNu}). Besides, the PIV measurements do not show any increase in velocity with $x$, confirming our hypothesis that the flow is at equilibrium, balanced between gravity acceleration and bed friction. This justifies to use the equilibrium solutions despite the small size of the experiment. We finally note that we were not able to observe any hydrodynamic structure such as those described in the discussion of the main text. This could be due to the depth-averaging of our PIV measurements.

Finally, in our experiments, the Reynolds number  $Re=U\,h/\nu$ ranges from $36$ to more than 400. These values 
exceed greatly the threshold of generation of Kapitza waves at the free surface $Re_c=5/(4\,\tan \theta)$ \cite{Liu1993,CharruBook}. For $\theta=39^\circ$, $Re_c \approx 1.5$. Kapitza waves, which are transverse waves, then propagate continuously on the film surface. These waves likely enhance the solute transport. As their velocity is typically $2\,U$, their dynamics is very fast compared to the erosion time scales. We thus neglect them, and only consider the time-averaged hydrodynamics to study the dissolution process. 

\newpage
\section{Eroded surface characterization.}
\label{surfchar}

 \begin{figure}[h!]
 \begin{center}
  \includegraphics[width=0.48\columnwidth]{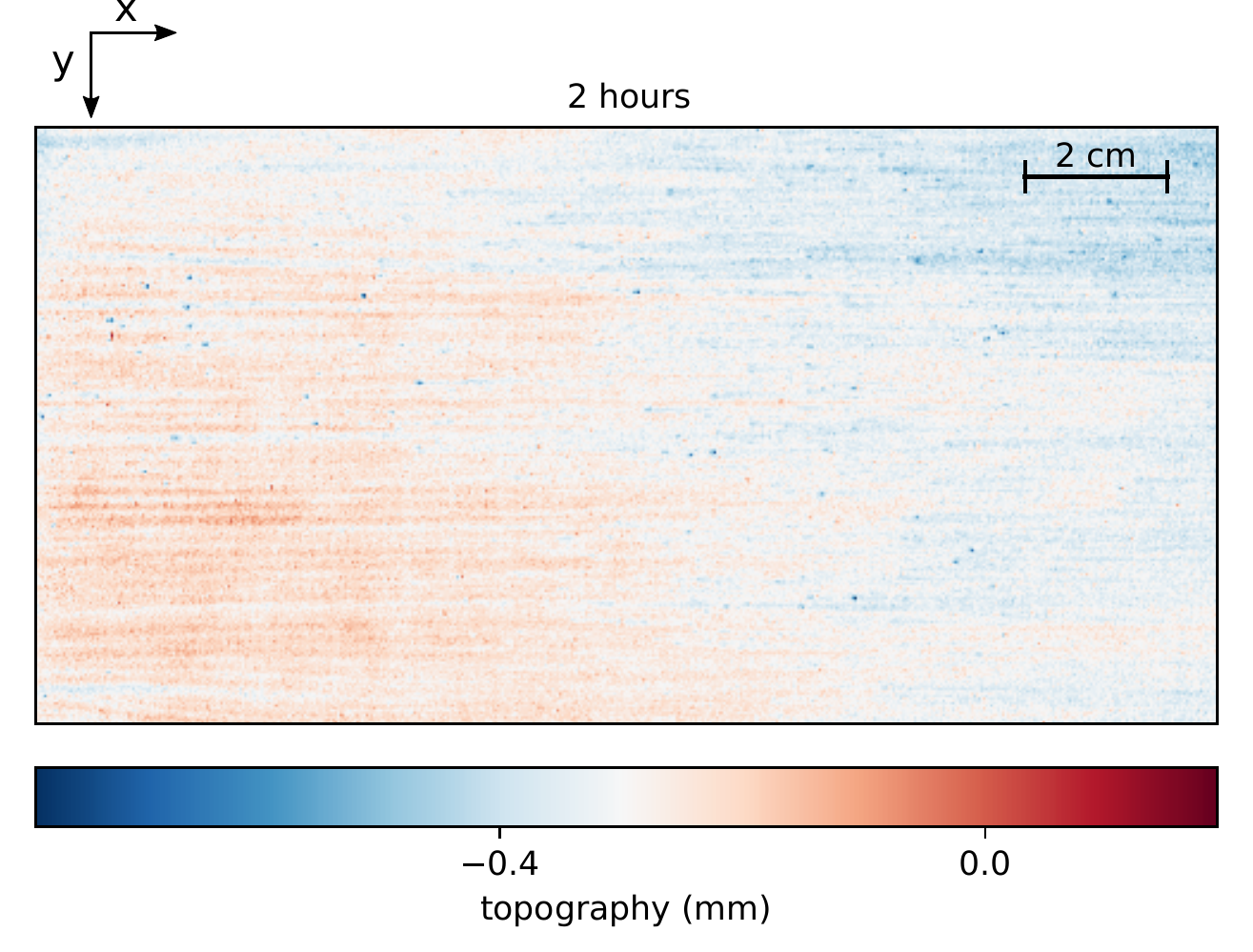} 
  \hfill
 \includegraphics[width=0.48\columnwidth]{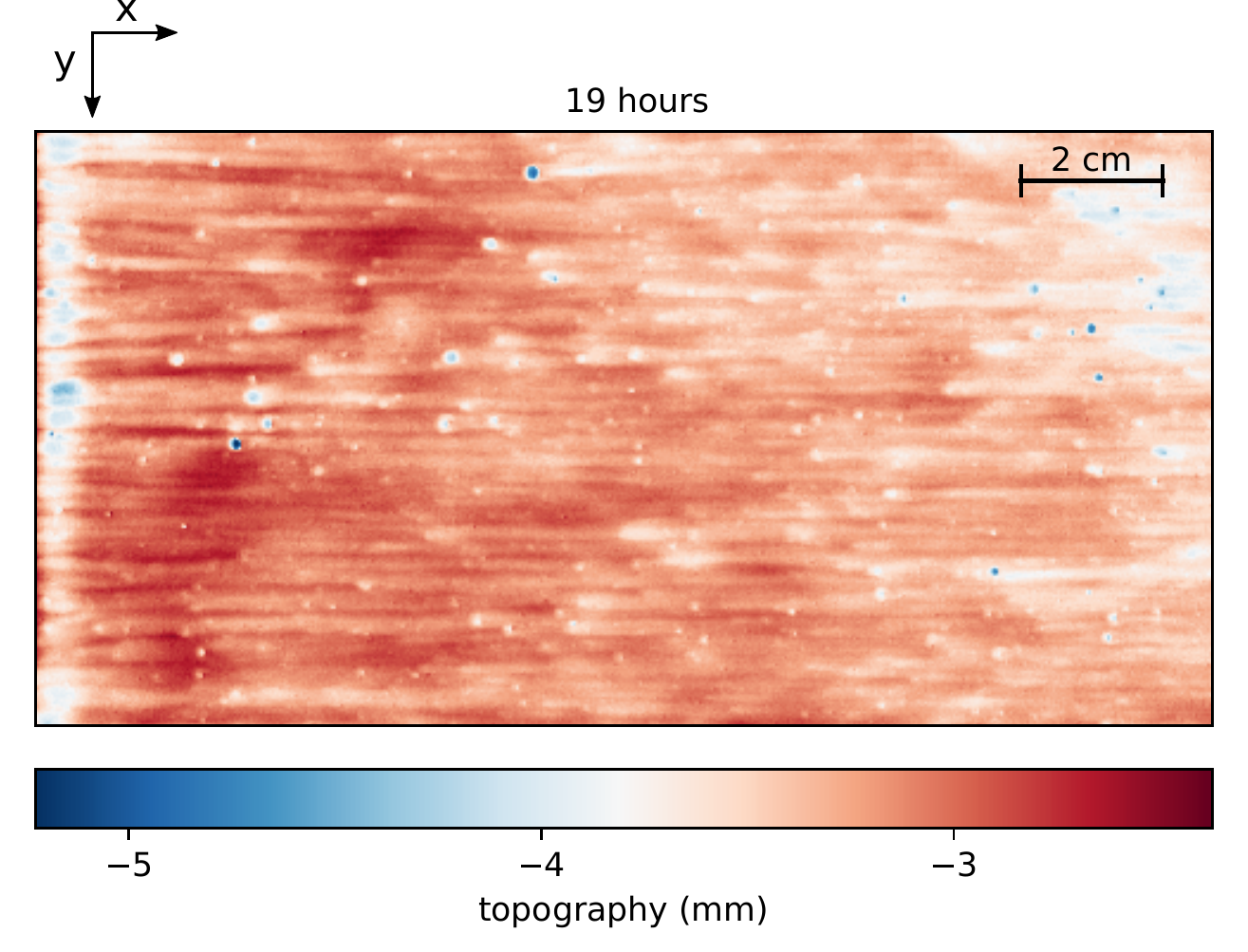}
\vspace{0.2cm}
 \caption{(a) Topography of the top surface of a plaster block inclined of  $39^\circ$ exposed to a fall film of flowrate $Q=2.8\,$L/min during $t=2\,$hours. We note a large scale variation, due to imperfections in the making of the initial flat surface or in the measurement. Thin dissolution groove are yet visible.  (b) Idem after $t=19$\,hours. The dissolution pattern remains longitudinal, but its amplitude increases associated with a widening of the grooves.}
    \label{figsup}
       \end{center}
 \end{figure}

\paragraph{Topography measurements}
The 3D topography of the eroded surface is measured with a laser scanner scanCONTROL 2900-100/BL (Micro-Epsilon 3D$\texttrademark$) associated to a motorized translation stage LTS300 Thorlabs$\texttrademark$. This device provides the elevation of the surface with an accuracy of $70\,\mu$m and a horizontal resolution of $0.2$\,mm. After interpolation, we obtain the maps of the topography $\eta(x,y,t)$ as a matrix.

This measurement is directly performed on the dried plaster blocks, after the experiment has been transiently stopped. In contrast, the pink salt blocks are translucent and the laser scanner cannot be used directly on the rock. 

Examples of 3D reconstructions of plaster eroded surface are plotted in Fig.~\ref{figsup}. {The tip of the block at large values of $x$ experiences a significant erosion rate. Consequently the length of the block decreases with time of typically $15$ mm in 50 hours. This observation is not studied here, because it corresponds to another hydrodynamic configuration. In order to compare the different states of one block in comparable situations, the topography is analyzed in a restricted $x$ interval smaller than the block length. For the block in Fig.~\ref{figsup}, the displayed length is $166$\, mm, which is smaller than the initial surface of the block subjected the flow, about $180$\,mm. Morever,} we note the presence of small circular cavities about $1$\,mm in diameter due to the presence of trapped bubbles in the plaster during its molding. They are not present initially on the top surface of the blocks, but become visible after an hour or so. Here, we must stress that we do not observe any similar defects on the salt blocks and on the alabaster plate, although we report a similar appearance of streamwise grooves. Therefore, we assume that these bubbles are not at the origin of the observed patterns, yet they induce an additional noise in the data processing. Moreover, large-scale structures can mask the small-scale dissolution patterns. Imperfections in the making of plaster blocks or in the horizontal leveling and increased erosion near the water injection can explain such large-scale topography.\\

\paragraph{Analysis of transverse profiles}
The analysis is performed on transverse profiles along the $y$-axis. The profiles are averaged over an interval spanning $4$\,mm along $x$, thus defining a transverse profile at a given value of $x$ and $t$: $\eta_{x,t} (y)=\langle \eta (x,y,t) \rangle_{[x-2,x+2]}$. To remove the large-scale trend irrelevant to the small-scale pattern formation, we subtract to each profile its fit by a polynomial of degree 3.\\

 \begin{figure}[h!]
 \begin{center}
  \includegraphics[width=0.48\columnwidth]{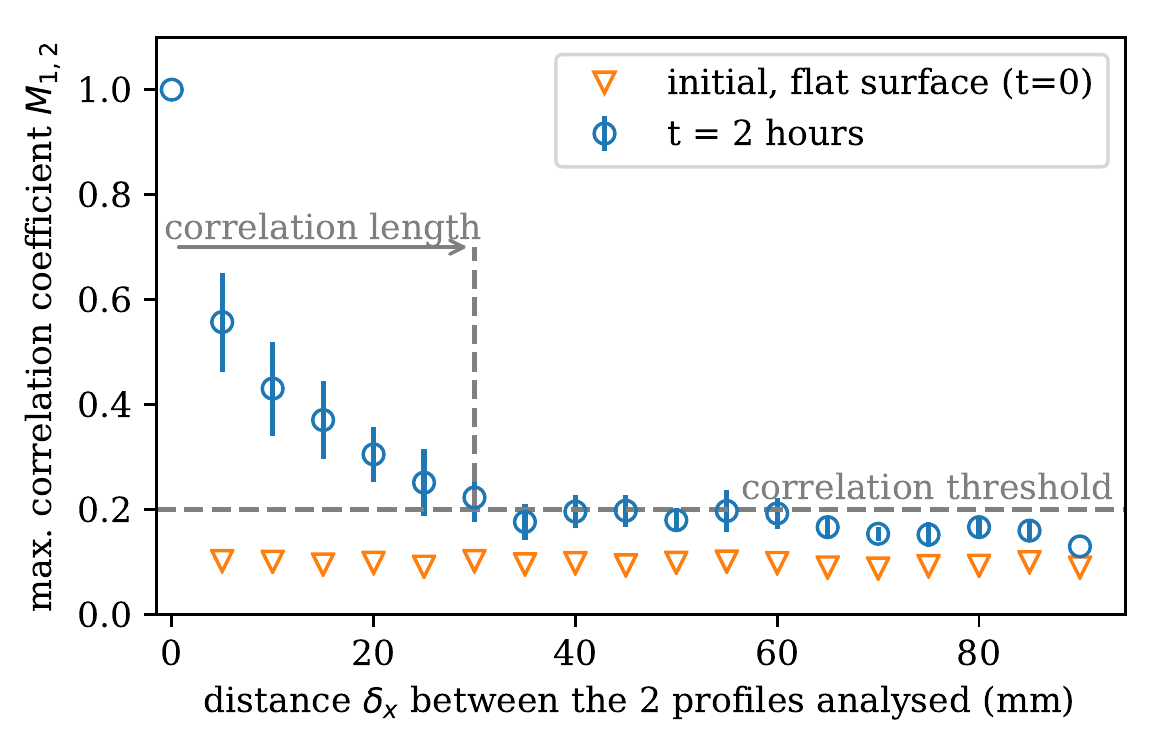}
  \caption{Correlation coefficient between two transverse profiles with respect to the distance along $x$ separating the profiles. Orange triangles: initial topography (flat and smooth surface), defining the correlation threshold at 0.2, that is two times the mean value of its correlation coefficients. Blue points: analysis of the topography after 2 hours, for $Q=2.8$\,L/min and $\theta = 39$\,deg. Above the correlation threshold, two profiles are considered to be correlated. This defines the correlation length as the first distance where the correlation coefficient is below the threshold. {Several distinct couples of profiles correspond to one separation distance $\delta_x$. The correlation coefficient is computed as the average on these couples and the error bars are the corresponding standard deviation.}}
    \label{figsup:correlation_coefficient}
    \end{center}
 \end{figure}

\newpage
\paragraph{Correlation length between transverse profiles}
Then, we demonstrate the streamwise character of the dissolution pattern by computing the correlation between transverse profiles. At a given time $t_0$, we compute the normalized cross-correlation $C_{1,2}$ between two transverse profiles $\eta_1$ and $\eta_2$ located at two different longitudinal positions $x_1$ and $x_2$:
\begin{equation}
C_{1,2}(\delta_y)=\dfrac{1}{\sigma_{\eta_1}\,\sigma_{\eta_2}}\, \int_{y_{min}}^{y_{max}} \, \eta_{1} (y) \times \eta_2 (y-\delta_y)   \, \mathrm{d}y \, ,
\end{equation}
where $\sigma_{\eta_1}$ and $\sigma_{\eta_2}$ are the standard deviations of $\eta_1$ and $\eta_2$ (see equation~(\ref{eq:sigma_eta}) below), and $\delta y$ is a shift along $y$ between the two profiles and ranges from 0 to the length of the profile (which is actually the width of the block). The correlation maximum $M_{1,2}$ of $C_{1,2}(\delta_y)$ then defines the correlation coefficient between the two profiles. The correlation coefficient of two identical profiles is 1, and decreases towards 0 as the correlation between the profiles decreases. Then, we study the dependency of $M_{1,2}$ as a function of the distance between the profiles $\delta_x=|x_1-x_2|$. By definition, $M_{1,2} (\delta_x=0)$ equals $1$. Then, $M_{1,2} (\delta_x)$ decreases with $\delta_x$, with fluctuations due to the experimental noise (blue points, Fig.~\ref{figsup:correlation_coefficient}). In contrast, the typical correlation coefficient at $t=0$ is independent of $\delta_x$ and fluctuates around a mean value of 0.1 (orange triangles). We arbitrarily define the correlation threshold at 0.2, that is two times the typical value of the coefficients at $t=0$. Above this threshold, we consider that two profiles are correlated. Then, we define a longitudinal correlation length $L_C$, which is the first $\delta_x$ position where $M_{1,2}<0.2$. We consider this correlation length as a fair mathematical representation of the average length of the grooves observed in our experiment. Fig~2 (d) of the main document shows its evolution during the experiment with $Q=2.8$\,L/min and $\theta = 39$\,deg, showing that it grows as the pattern forms and develops. This figure has been obtained for the experiment $U=0.84\,$m/s, but the other experiments give very similar results.

Finally, it is to note that the location $\delta_{y,\mathrm{max}}$ of $M_{1,2}$ along $\delta_y$ indicates a possible lateral shift of the pattern. In our experiments, when $M_{1,2}$ is above the correlation threshold, we almost always obtain $\delta_{y,\mathrm{max}}=0$, which gives another clue that the dissolution grooves are streamwise.\\

\begin{figure}
 \begin{center}
  \includegraphics[width=.48\columnwidth]{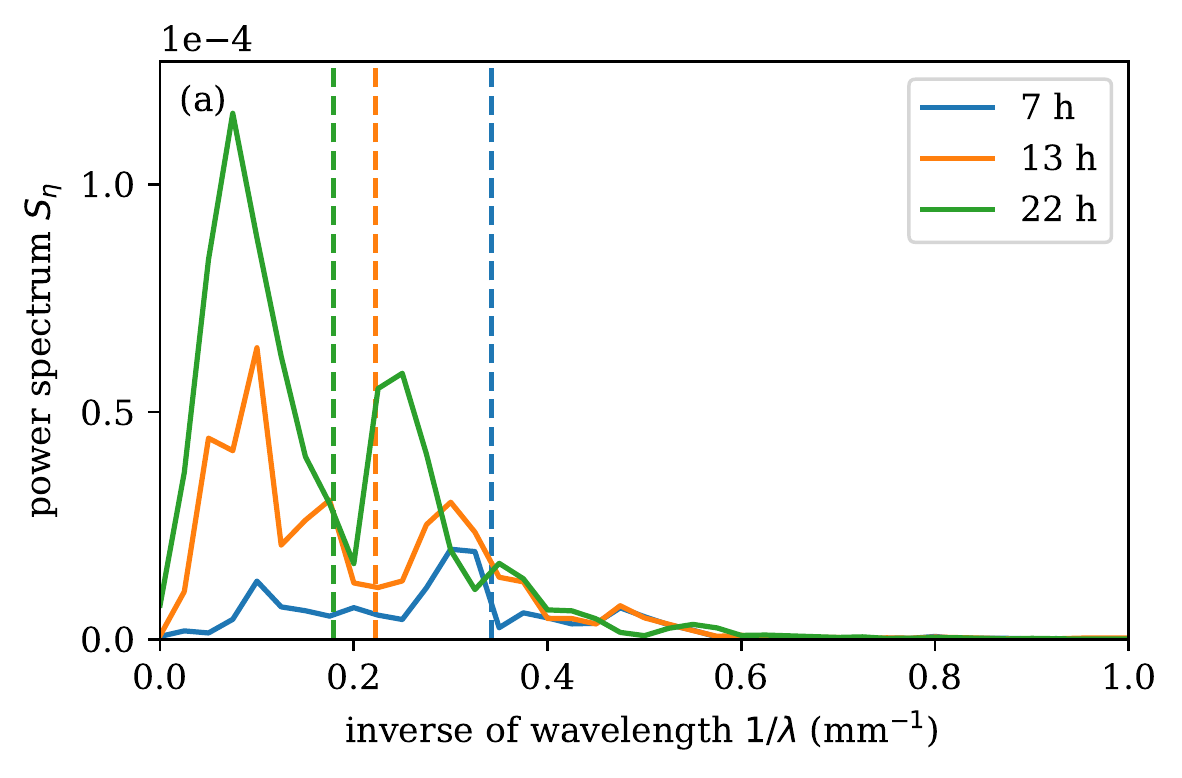} 
  \hfill
  \includegraphics[width=0.48\columnwidth]{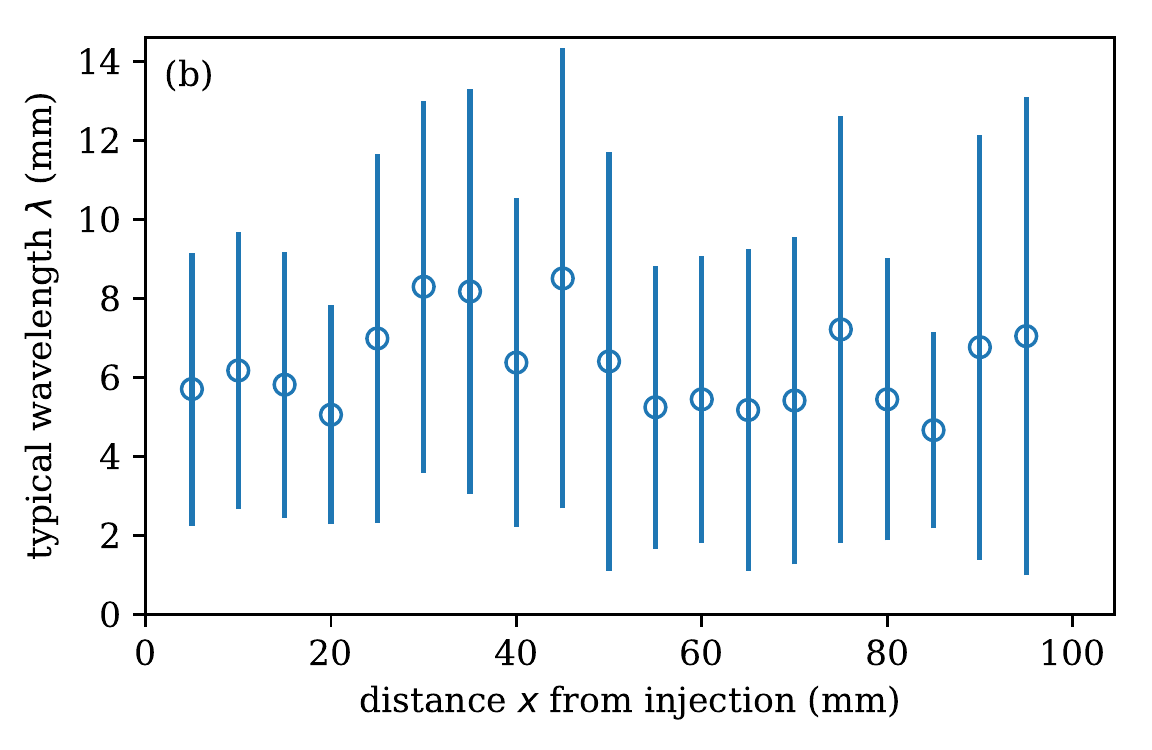}
  \end{center}
 \caption{(a) Space Fourier power spectrum $S_\eta$ of the profiles along $y$ $\eta (x^*,y,t^*)$ with $x^*=20\,$mm and $t=7$, $13$ and $22$ hours as a function of the inverse of the wavelength ($1/\lambda$). Dashed vertical lines point to the typical inverse wavelength $\lambda_m^{-1}$, computed as the weighted average of the spectrum. (b) Typical grooves wavelength $\lambda_m$ with respect to the distance $x$ from the water injector, after a time $t=22$~hours of exposure to the flow. {The error bars are estimated by computing the second moment of the spectrum as a standard deviation}. Both graphics are made for the experiment with $Q=2.8$~L/min and $\theta=39$~deg ($U=0.84\,$m/s). }
 \label{FigSp}
 \end{figure}

\paragraph{Typical wavelength of the pattern}
The estimation of the typical width of the grooves presents some difficulties. The pictures {of the eroded surfaces (see \S IV)} show a characteristic width of order one millimeter, but not a well-defined periodicity. We propose a spatial Fourier analysis of the transverse profiles. For each time step $t$, we compute the spatial power spectrum $S_\eta (1/\lambda)$ of profiles $\eta_{x,t}(y)$ for various positions $x$ of the profile, $x$ varying from 5\,mm to the length of the block and with a step of 5\,mm. As an example, we plotted in Fig.~\ref{FigSp} (a) the spectrum $S_\eta (1/\lambda)$ as a function of the inverse of the wavelength, calculated at the position $x=20\,$mm, for $t=7$, $13$ and $22$~hours. A quite wide peak at scales $1/\lambda<1$\,mm$^{-1}$ denotes the appearance of the pattern. As time increases, the maximum of the spectrum is displaced to larger scales. A secondary peak appears at a wavelength two times larger than the maximum, and is visible for long enough times.

In order to obtain a systematic and robust measurement of the characteristic grooves width, a typical wavelength $\lambda_m$ is extracted from the spectrum by an average weighted by the spectrum:
\begin{equation}
\lambda_m^{-1}=\dfrac{\int_0^\infty \lambda^{-1} \, S_\eta(1/\lambda) \,\mathrm{d} (1/\lambda)}{\int_0^\infty  S_\eta(1/\lambda) \,\mathrm{d} (1/\lambda)} \, .
\end{equation}
Thus, we obtain a typical wavelength with respect to the position $x$ for each time step. Fig.~\ref{FigSp} (b) shows that $\lambda_m$ does not depend on $x$, with a typical example taken at $t=22$~hours, $Q=2.8$~L/min and $\theta=39$~deg. It fluctuates around a mean value that is constant along the whole length of the block, demonstrating again the streamwise character of the pattern.

In order to reduce the experimental noise inherent to our measurements, we evaluate the typical width $\lambda_m$ of the grooves at a time $t$ as the mean value of $\lambda_m (x)$, averaged over the length of the block. The time evolution of $\lambda_m$ such computed is plotted in Fig.~3 (a) of the main document and shows a regular evolution. Finally, other methods to estimate the grooves typical width (autocorrelation and average crest-to-crest distance) provide the same results at a given multiplicative factor of order one.\\

\begin{figure}
 \begin{center}
  \includegraphics[width=0.48\columnwidth]{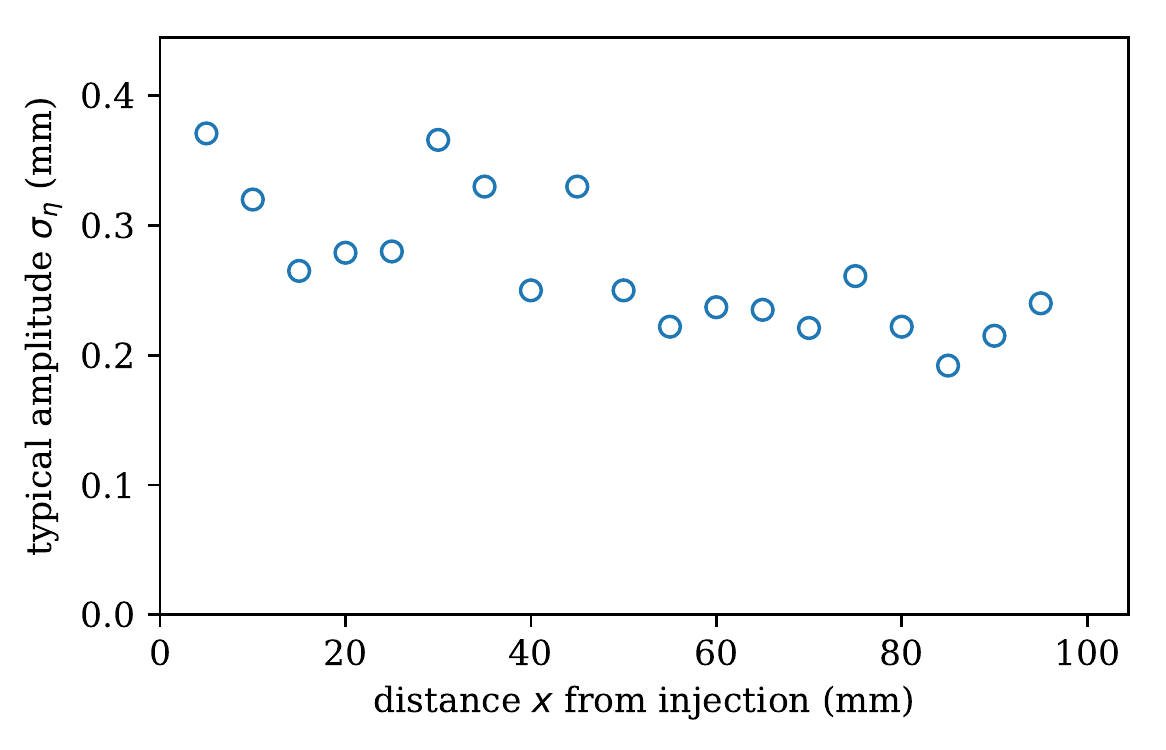}
  \caption{Pattern amplitude with respect to the distance $x$ from the water injection, at $t=19$~hours, for $Q=2.8$~L/min and $\theta=39$~deg ($U=0.84\,$m/s).}
    \label{figsup:amplitude_VS_position}
    \end{center}
 \end{figure}

\paragraph{Typical amplitude of the pattern}
We evaluate the pattern amplitude by the standard deviation of the transverse profiles: 
\begin{equation} \label{eq:sigma_eta}
\sigma_{\eta} (x,t)=\left[\langle \eta_{x,t}^2\rangle_y - (\langle \eta_{x,t} \rangle_y)^2 \right]^{1/2}\, ,
\end{equation}
which is in fact the measure of a typical rugosity. Thus, at each time step, we measure a typical amplitude $\sigma_\eta (x)$ for various positions ranging from $x=5$~mm to the rock length. Figure~\ref{figsup:amplitude_VS_position} shows that $\sigma_\eta (x)$ slightly decreases with $x$, most certainly due to the fact that the rock erosion is larger as $x \to 0$. Yet, to be consistent with our measurement of $\lambda_m$, which is averaged over $x$, we estimate a single typical pattern amplitude $\sigma_\eta$ by averaging $\sigma_\eta (x)$ over the block length.

 \newpage
 \section{Experimental runs}
\label{expruns} 
\begin{figure}[h!]
 \begin{center}
    \includegraphics[width=1\columnwidth]{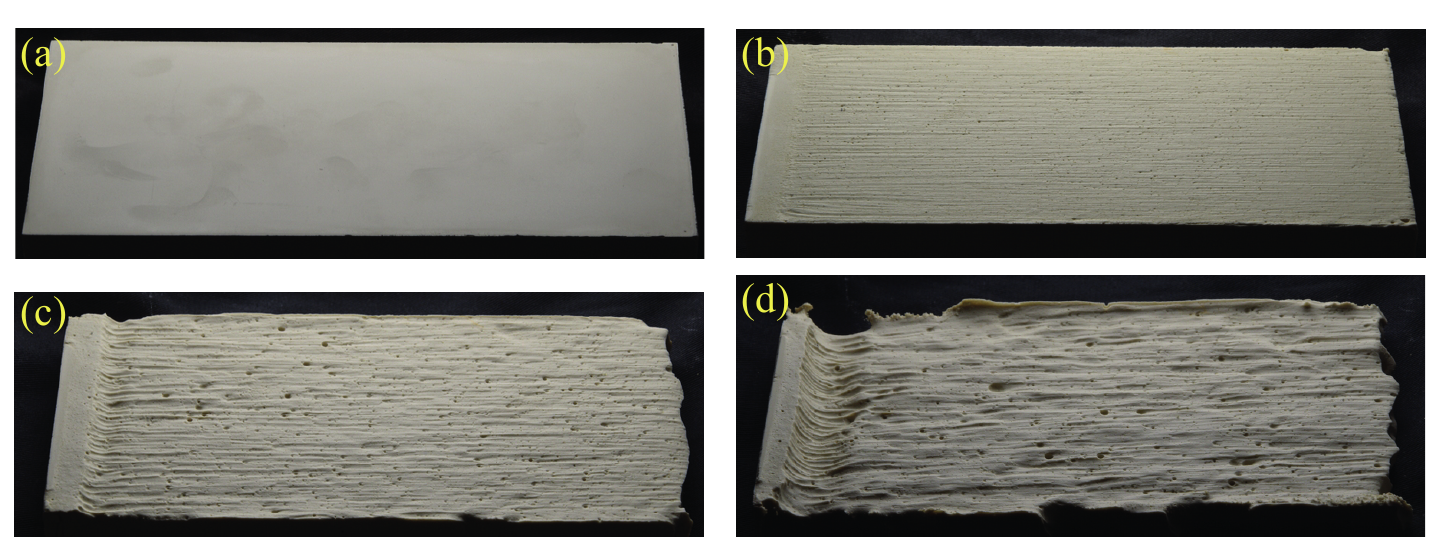} \\
   \end{center}
 \caption{Consecutive pictures of a same plaster block, in the case $Q=2.8\,$L/min, $\theta=39^\circ$ and $U=0.84$\,m/s at different time steps. This experiment is described in details in the main document. (a) Initial state $t=0$ hours. (b) $t=2$ hours. (c) $t=19$ hours. (d) $t=45$ hours. The initial dimensions of the block are $200\times 100 \times 25\,$mm.}
 \label{Figrun4}
 \vspace{2cm}
 \end{figure}

\begin{figure}[h!]
 \begin{center}
    \includegraphics[width=1\columnwidth]{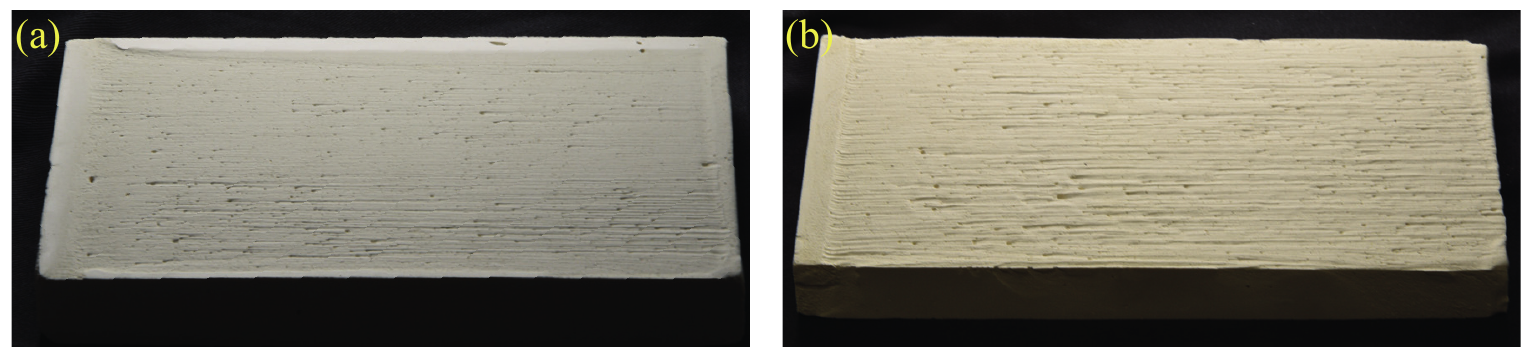} \\
   \end{center}
 \caption{Emergence of small grooves at moderate flowrate. (a) $Q=0.22\,$L/min, $\theta=39^\circ$ and $U=0.36$\,m/s.  $t=4$ hours. (b) $Q=0.87\,$L/min, $\theta=39^\circ$ and $U=0.57$\,m/s.  $t=8$ hours.}
 \label{Figrun3}
 \end{figure}

\begin{figure}[h!]
 \begin{center}
    \includegraphics[width=1\columnwidth]{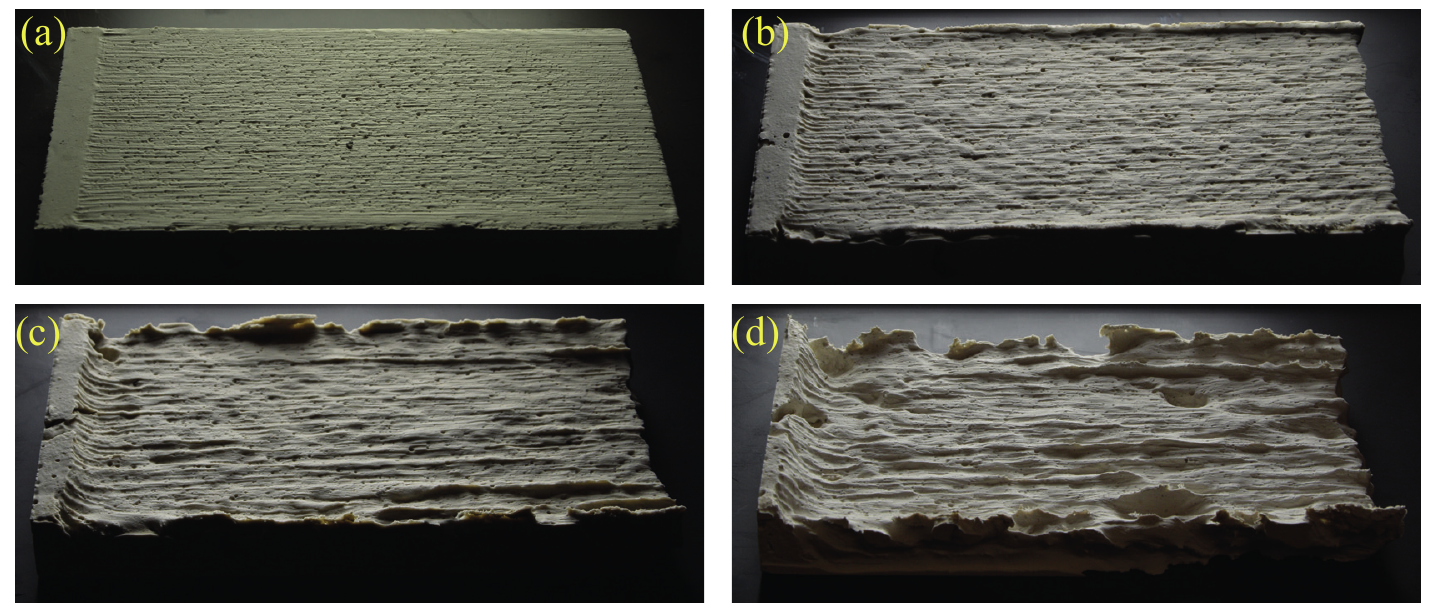} \\
   \end{center}
 \caption{Consecutive pictures of a same plaster block, in the case $Q=1.91\,$L/min, $\theta=66^\circ$ and $U=0.84$\,m/s at different time steps.  (a)  $t=2$ hours. (b) $t=16$ hours. (c) $t=37$ hours. (d) $t=58$ hours. The initial dimensions of the block are $200\times 100 \times 25\,$mm.}
 \label{Figrun6}
  \vspace{2cm}
 \end{figure}

\begin{figure}[h!]
 \begin{center}
    \includegraphics[width=1\columnwidth]{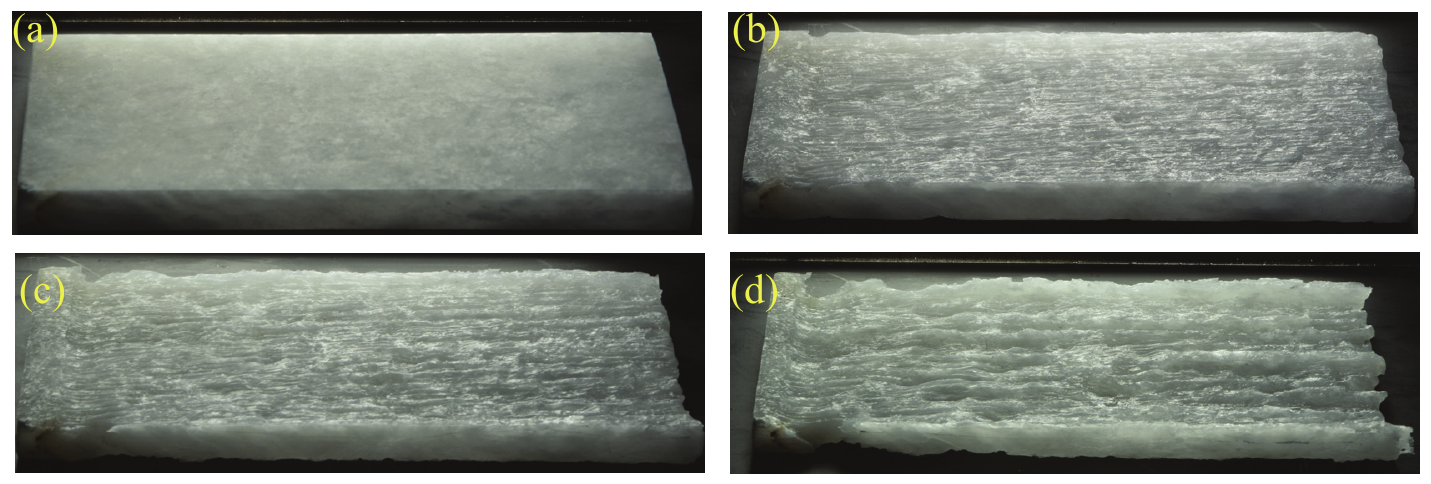} \\
   \end{center}
 \caption{Pictures of the unique experiment performed with a polished alabaster plate. (a) Initial state. Dimensions $200 \times 100 \times 15$ mm. (b) $Q=0.97\,$L/min, $\theta=25^\circ$ and $U=0.51$\,m/s at $t=17$ h. (c)  $Q=0.97\,$L/min, $\theta=25^\circ$ and $U=0.51$\,m/s at $t=33$ h. (d) $Q=0.97\,$L/min, $\theta=25^\circ$ and $U=0.51$\,m/s at $t=60$ h. We thank the \textit{Atelier Alain Ellouz} ( https://atelier-alain-ellouz.com/~) for giving us a free Alabaster plate to test in our experiment. Note that the patterns observed in this experiment are very similar to those observed on plaster, another kind of gypsum, although \textit{i)} their surface texture differ and \textit{ii)} bubbles are absent in alabaster.}
 \label{Figrunalabaster}
 \end{figure}

\begin{figure}[h!]
 \begin{center}
    \includegraphics[width=1\columnwidth]{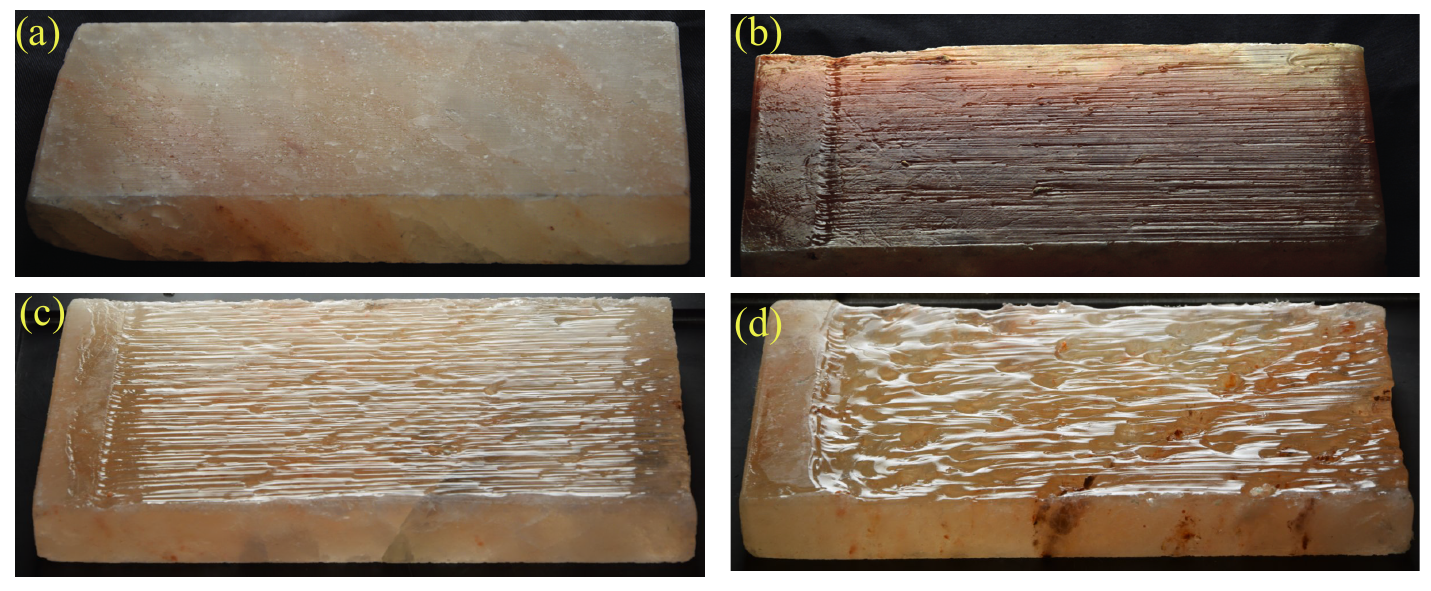} \\
   \end{center}
 \caption{Pictures of experiments performed with Himalaya pink salt blocks. (a) Initial state. (b) $Q=0.79\,$L/min, $\theta=36^\circ$ and $U=0.55$\,m/s at $t=1$ min. (c)  $Q=1.84\,$L/min, $\theta=47^\circ$ and $U=0.79$\,m/s at $t=2$ min. (d) $Q=1.84\,$L/min, $\theta=47^\circ$ and $U=0.79$\,m/s at $t=5$ min. We thank Jean-Gabriel Pichon, an undergraduate intern, who performed these two last runs.}
 \label{Figrunsel}
 \end{figure}

In this section, we provide
\begin{itemize}
	\item a table with the experimental parameters (table~\ref{tab:exp_params}),
	\item some pictures of plaster blocks taken after various times of exposure to the flow, and for various experimental parameters (figs.~\ref{Figrun4}, \ref{Figrun3} and \ref{Figrun6}),
		\item some pictures of the unique experiment performed on an Alabaster plate (fig.~\ref{Figrunalabaster}),
	\item and some pictures of experiments performed on salt blocks (fig.~\ref{Figrunsel}).
\end{itemize}

\begin{table}[h!]
\begin{tabular}{|c|c|c|c|c|c|c|c|c|}
\hline 
& Flow rate $Q$ & Inclination & Flow velocity  & Film thickness  & Reynolds number & First time step  & Largest exposure &{Erosion rate}  \\ 
& (L/min)  &  $\theta$ ($\circ$)  & $U$ (m/s) &  $h$ (mm) &  $Re=h\,U/\nu$ & with rills (hours) & time (hours) & {$\times 10^{-8}$ (m/s)} \\ 
\hline 
1 ($\square$) & 0.22 & 39 & 0.36 & 0.101 & 36 & 1 & 4 & 2.4 \\ 
\hline 
2  ($\triangledown$) & 0.86 & 39 & 0.57 & 0.255 & 143 & 0.5  & 11 & 3.84 \\ 
\hline 
3 ($\circ$) & 2.8 & 39 & 0.84 & 0.556 & 466 & 0.5  & 55 & 5.15 \\ 
\hline 
4 ($+$) & 1.9 & 66 & 0.84 & 0.380 & 318 & 1 & 71&6.49 \\ 
\hline 
5 ($\star$) & 0.93 & 25 & 0.51 & 0.305 & 155 & 1 & 13.5& 4.62\\ 
\hline 
\end{tabular} 
\caption{Parameters of the five plaster experiments quantitatively analyzed. }
\label{tab:exp_params}
 \vspace{1cm}
\end{table}

\newpage
~
\newpage
~
\newpage

  \section{Linear stability analysis of the dissolving bed in the transverse plane.}
  \label{linstab}
  
 
 
In order to explain the emergence of dissolution patterns, we propose a simplified linear stability analysis of the bed subjected to the flow. The erosion rate of a soluble rock depends on the concentration of solute at the boundary between the rock and water. In order to predict the erosion pattern, we must therefore know the concentration field of solute in the water flow. For that purpose, we analyze in stationary regime, how the rock is dissolved into water, and then how it is transported by the water flow. We use the same set of equations than in Philippi et al.~\cite{Philippi2019}, which describes the dissolution process and its coupling to the flow.

\subsection{Advection-diffusion equation}

The solute is transported by advection and by diffusion. Thus, we must solve the stationary advection-diffusion equation, which reads
\begin{equation} \label{eq:full_adv-diff}
	\vec{u} \cdot \vec{\nabla}c(x,y,z) = D\, \Delta c(x,y,z) \, ,
\end{equation}
where $\Delta$ is the laplacian operator and $D$ is the solute diffusion coefficient and where we have assumed that the system is in steady state. {For gypsum~\cite{Colombani2007} and salt~\cite{Handbook} in water}, $D\approx 10^{-9}\,$m$^2$\,s$^{-1}$. If one assumes that there is no velocity in the $y-$ and $z-$direction ($\vec{u} = u(x,y,z)\, \vec{e_x}$), this equation becomes:
\begin{equation} \label{eq:adv-diff}
	u\, \p_x c = D\, \left(\p_{x,x} c + \p_{y,y} c + \p_{z,z} c \right) \, .
\end{equation}
This equation must be supplemented by 3 boundary conditions:
\begin{equation} \label{eq:adv-diff_BCs}
\left\{
\begin{array}{lcl}
	c = 0 & \mathrm{in} & x=0 \\
	\p_z c = 0 & \mathrm{in} & z=h \\
	\alpha\, (c-c_{\mathrm{sat}}) = D\, \p_z c & \mathrm{in} & z=\eta \, ,
\end{array}
\right.
\end{equation}
where $\alpha$ is the dissolution coefficient. For gypsum in water, $\alpha \approx 2.7 \cdot 10^{-6}\,$m\,s$^{-1}$~\cite{Colombani2007} and for salt in water, $\alpha \approx 5.0 \cdot 10^{-4}\,$m\,s$^{-1}$~\cite{Alkattan}. The boundary condition along the bottom is important because it determines the dissolution rate $\alpha\, (c-c_{\mathrm{sat}})$, and thus the erosion rate.

\subsection{Advection-diffusion equation at order 0} 
The solution at order 0 to our problem, or {\it base state}, can be tackled analytically provided several approximations:
\begin{itemize}

	\item We assume that the velocity field can be approximated by its average value $U$ in the advection-diffusion equation~(\ref{eq:adv-diff}). This hypothesis of homogeneous flow may appear crude, as it neglects the velocity decrease close to the bed, where the solute concentration and thus the solute transport are the highest. Yet, we note that a similar analysis using the Nusselt velocity profile (half parabolic) provide qualitatively similar results (Private communication of Carlo Camporeale and Matteo Bertagni). Moreover, the flow characterization \S~\ref{flowcha} shows that the films in the conditions of our experiment belong to a regime of a wall-induced turbulence. Therefore, with a simplified hydrodynamics, we aim here to identify qualitatively the mechanisms at the origin of dissolution patterns and we do not expect a quantitative agreement.
	
		\item We assume that the bed elevation does not depend on $y$. $\eta = \eta_0 (x,t)$ with $\eta_0 (x, t=0)=0$. We further assume that $\eta_0 \ll h$, such that the variation of the bed elevation does not perturb the velocity field.
	
	\item We neglect possible density stratification and gravity effects, due to the increase of the local fluid density with the concentration. For gypsum, at the saturation concentration $c_{sat} \approx 2.58\,$kg.m$^{-3}$, the maximal fluid density is $\rho_{m} \approx 1002 \,$kg.m$^{-3}$, whereas for salt, $c_{sat} \approx 315\,$kg.m$^{-3}$, the maximal fluid density is $\rho_{m} \approx 1200 \,$kg.m$^{-3}$. These effects are thus likely incidental for gypsum and plaster, but they may play a role to determine the precise erosion rate for salt. In this last case, $U^2/h$ remains large compared to $g\,(\rho_m-\rho_0)/\rho_0$.
	
\item The distance $D/U \approx 2\,10^{-9}$ m is very small in the conditions of our experiments. We can assume that the Péclet number built along the coordinate $x$, $Pe=U\,x/D$ is always very large, such that we neglect the diffusion in $x$. We leave out also the diffusion in $y$, which could be taken into account in a more elaborate model. This simplifies equation~(\ref{eq:adv-diff}) into
		\begin{equation} \label{eq:adv-diff_O1}
			U\,\p_x c_0(x,z) = D\, \p_{z,z} c_0(x,z) \, ,
		\end{equation}
\end{itemize}
which is now similar to the 1D-diffusion equation, where time has been replaced by $x$. In order to solve this equation, we will make two additional approximations regarding the boundary conditions.

\subsection{Dissolution of the bed infinitely fast compared to diffusion} \label{sec:errate_0_c=csat}

First, we assume that the Damköhler number $\Da = \alpha\,h/D$ is very large compared to one (in fact, $\Da \approx200$ for salt and $\Da\approx 1$ for plaster). The boundary condition in $z=\eta$ is therefore simplified into
\begin{equation} \label{eq:BC_large_dahmkohler}
	c_0 = c_{\mathrm{sat}} ~~ \mathrm{in} ~~ z=\eta_0 \approx 0 \,.
\end{equation}

Second, the upwards diffusion is slow compared to the downstream advection. Indeed, the concentration front would reach the free surface for a length  $L=(h^2\,U)/D \approx 5$ m, which is much larger than our experimental blocks. This implies that the solute does not have the time to diffuse up to the fluid surface, which can therefore be seen as infinitely far from the layer of fluid that is loaded with solute. With this assumption, one can consider the fluid domain as semi-infinite, and the condition in $z=h$ becomes mathematically:
\begin{equation}
	c_0 = 0 ~~ \mathrm{for} ~~ z \to \infty \,.
\end{equation}

A solution to this problem in the limit of a semi-infinite medium reads~\cite{crank1979mathematics}:
\begin{equation} \label{eq:c0_dissolution_kinetics_infinite}
	c_0 (x,z) = c_{\mathrm{sat}} \, \left[ 1- \erf \, \left( \dfrac{z\, \sqrt{U}}{2\,\sqrt{D\, x}} \right) \right] \, ,
\end{equation}
where erf is the error function:
\begin{equation}
	\erf \, (z) = \dfrac{2}{\sqrt{\pi}} \, \int_0^z \exp (-\eta^2) \, \mathrm{d}\eta \, .
\end{equation}

The erosion rate $V_{\mathrm{er},0}$ of the plaster block is then given by the diffusive flux of solute in $z=\eta$, divided by the density of the dissolving solid $\rho_{rock}$. In units of a velocity,
\begin{equation} \label{eq:erosion_rate}
	V_{\mathrm{er,}0} (x) = \p_t \eta_0 (x, t) = \dfrac{D}{\rho_{\mathrm{rock}}} \, \p_z c_0 \,  (x, z=\eta_0) = 
			- \dfrac{ c_{\mathrm{sat}}\, \sqrt{U\, D}}{\rho_{\mathrm{rock}} \, \sqrt{\pi \, x}} \, .
\end{equation}

Here we can notice that, according to this model, the erosion velocity increases with the flow velocity and decreases with the distance to the water injection, which conforms to physical intuition. We also notice that the negative sign only indicates that the erosion is in the direction opposite to the upwards coordinate $z$. By integrating this expression over the length of the block, we obtain the average erosion velocity of the salt block: 
\begin{equation} \label{eq:erosion_rate_prediction}
\langle V_{\mathrm{er,}0} \rangle = - \dfrac{2\, c_{\mathrm{sat}}\, \sqrt{U\, D}}{\rho_{\mathrm{rock}} \, \sqrt{\pi \, L}} \, .
\end{equation}
We measured the density of the plaster blocks $\rho_{\mathrm{rock}}= \rho_{\mathrm{plaster}}\approx 1200 $ kg.m$^{-3}$. Colombani and Bert~\cite{Colombani2007} measured for gypsum $c_{sat}= 2.58 $ kg.m$^{-3}$ and $D=1.0\,10^{-9}$\,m$^2$\,s$^{-1}$. We predict thus an average erosion rate $\langle V_{\mathrm{er,}0} \rangle \approx 2.2 \cdot 10^{-7}$ m/s for the run $3$ ($U=0.84$m s$^{-1}$. Experimentally, the slope of the curve $\langle - \eta (t) \rangle$ as a function of $t$ (Fig.~2 of the main text) gives a constant erosion velocity $\langle V_{\mathrm{er,}0} \rangle \approx  5 \cdot 10^{-8}$ m/s. This simple model provides the correct order of magnitude. This comparison justifies a model relying on the diffusive solute transport perpendicular to the dissolving surface. However, a better modeling of the concentration and velocity boundary layers seems required to reach a quantitative agreement.

\subsection{Growth of a small perturbation}

We now assume that the bed elevation is slightly perturbed:
\begin{equation}
	\eta (x,y,t) = \eta_0(x,t) + \e \, \eta_1(y,t) \, ,
\end{equation}
with $\e \ll 1$.  This slight perturbation {\it a priori} perturbs both the flow velocity and the concentration fields. Yet, the Péclet number is much larger to 1, such that we assume that the concentration perturbation is negligible compared to the velocity perturbation. With this hypothesis, the forthcoming linear stability analysis can be performed in a plane transverse to the flow.

According to Chézy's law, the flow velocity is proportional to the square root of the flow height:
\begin{equation} \label{eq:linear_expansion}
\arraycolsep=1.8pt\def\arraystretch{2}
\begin{array}{ll}
	U = U_0 + \e \, U_1 & = \dfrac{1}{C_f^{1/2}} \, \sqrt{g\, \sin \theta\, (h - \eta)} \\
	  &  = \dfrac{1}{C_f^{1/2}} \, \sqrt{g\, \sin \theta\, (h_0 - \e \eta_1)} \\
	  &  \approx \dfrac{1}{C_f^{1/2}} \, \sqrt{g\, \sin \theta\, h_0} -\e \, \dfrac{1}{2\, C_f^{1/2}} \, \sqrt{\dfrac{g\, \sin \theta}{h_0}} \, \eta_1 \, .
\end{array}
\end{equation}
A linear expansion of the square root in equation~(\ref{eq:linear_expansion}) thus relates the velocity perturbation $U_1$ proportionnally to the bed perturbation $\eta_1$:
\begin{equation} \label{eq:velocity_expansion}
	U_1 = - \dfrac{1}{2\, C_f^{1/2}} \, \sqrt{\dfrac{g\, \sin \theta}{h_0}} \, \eta_1
			= - \dfrac{U_0}{2}\, \dfrac{\eta_1}{h_0} \, .
\end{equation}
The negative sign in equation~(\ref{eq:velocity_expansion}) implies that the flow velocity is larger above the troughs of the bed perturbation than above its crests. As the bed erosion increases with the flow velocity, this explains qualitatively why a bed perturbation will spontaneously amplify.

Formally, we call equation~(\ref{eq:erosion_rate}) to relate the erosion rate to the flow velocity:
\begin{equation}
	\p_t \eta (x, t) = \p_t \eta_0 (x, t) + \e\, \p_t \eta_1 (x, t) =
			- \dfrac{ c_{\mathrm{sat}}\, \sqrt{ D}}{\rho_{\mathrm{rock}} \, \sqrt{\pi \, x}} \, \sqrt{U_0 + \e \, U_1} \, .
\end{equation}
A linear expansion now gives
\begin{equation}
	\p_t \eta_1 (x, t) =
			- \dfrac{ c_{\mathrm{sat}}\, \sqrt{ D}}{2\, \rho_{\mathrm{rock}} \, \sqrt{\pi \, U_0 \, x}} \, U_1
			= \dfrac{ c_{\mathrm{sat}}\, \sqrt{U_0\, D}}{4\, \rho_{\mathrm{rock}} \, \sqrt{\pi  \, x}} \,\dfrac{\eta_1}{h_0}
			= \gamma\, \eta_1
\end{equation}
where
\begin{equation} \label{eq:growth_rate_prediction}
	\gamma = \dfrac{ c_{\mathrm{sat}}\, \sqrt{U_0\, D}}{4\, \rho_{\mathrm{rock}}\, h_0 \, \sqrt{\pi  \, x}}
\end{equation}
is the growth rate of the perturbation in this simple model. This value depends on the longitudinal coordinate $x$. As the experiments display streamwise patterns, we compute the average growth rate along $x$, by integrating along $x$ between $0$ and $L$ and dividing by $L$ with $L$ the block length ($0.1$ m), which reads:
\begin{equation} \label{eq:growth_rate_prediction2}
	\gamma_m = \dfrac{ c_{\mathrm{sat}}\, \sqrt{U_0\, D}}{2\, \rho_{\mathrm{rock}}\, h_0 \, \sqrt{\pi  \, L}}
\end{equation}

As it is positive and does not depend on the wavenumber, the perturbation is supposed to grow exponentially in time for all wavelengths. Any perturbation of the bed is thus amplified. However, with the hypotheses of the model, the growth rate does not depend on a transverse wavenumber $k_y$. This stability analysis does not explain the emergence of a pattern with a defined wavelength as all wavelengths are unstable. By incorporating the solute diffusion along the $y$ axis or by modeling the influence of the bed shape on the flow in viscous regime~\cite{Abramian2019}, one could show that the growth rate is damped at small wavelengths. However, in the absence of any mechanism decreasing the growth rate at large scales, the linear stability analysis of the bed cannot select any wavelength, as observed in the experiments.\\

	\begin{figure}[h]
	\centering
	  \includegraphics[width=.48\linewidth]{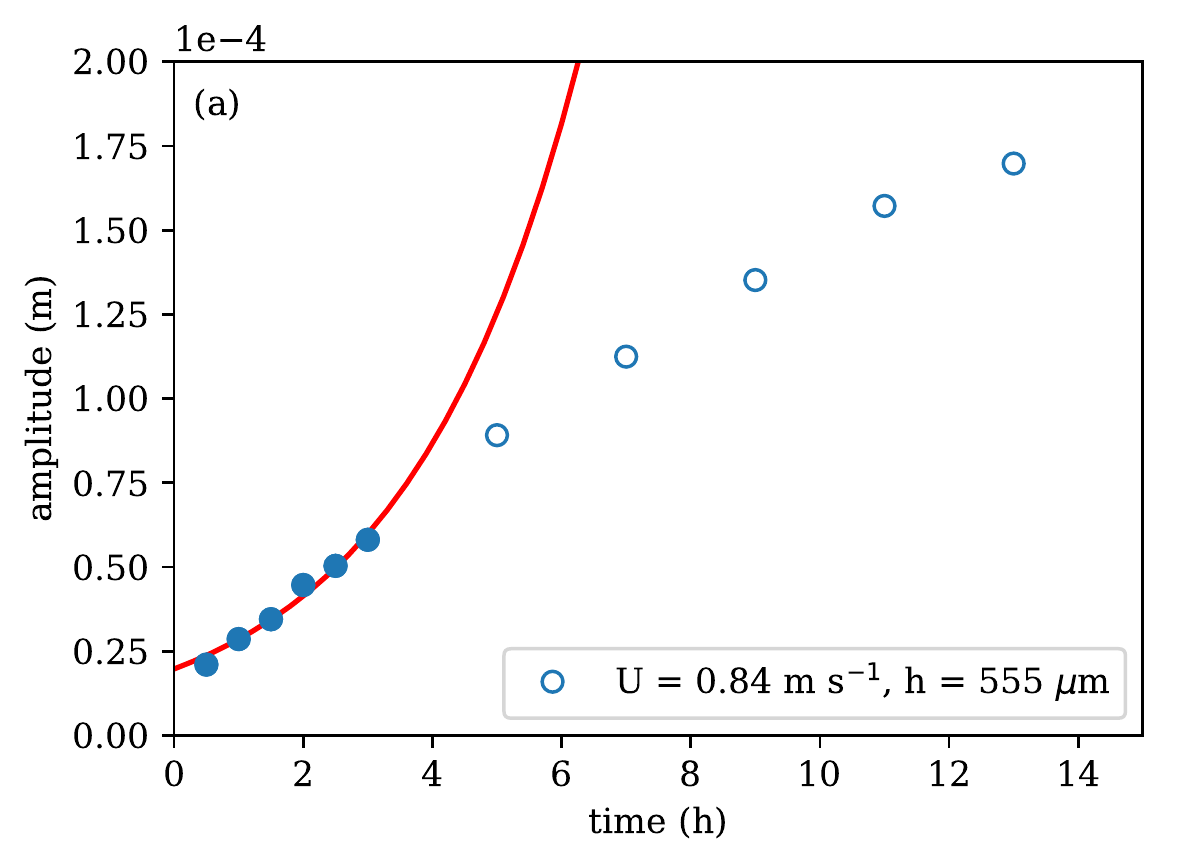}  
	    \label{subfig:WGR_h}
	    \includegraphics[width=.48\linewidth]{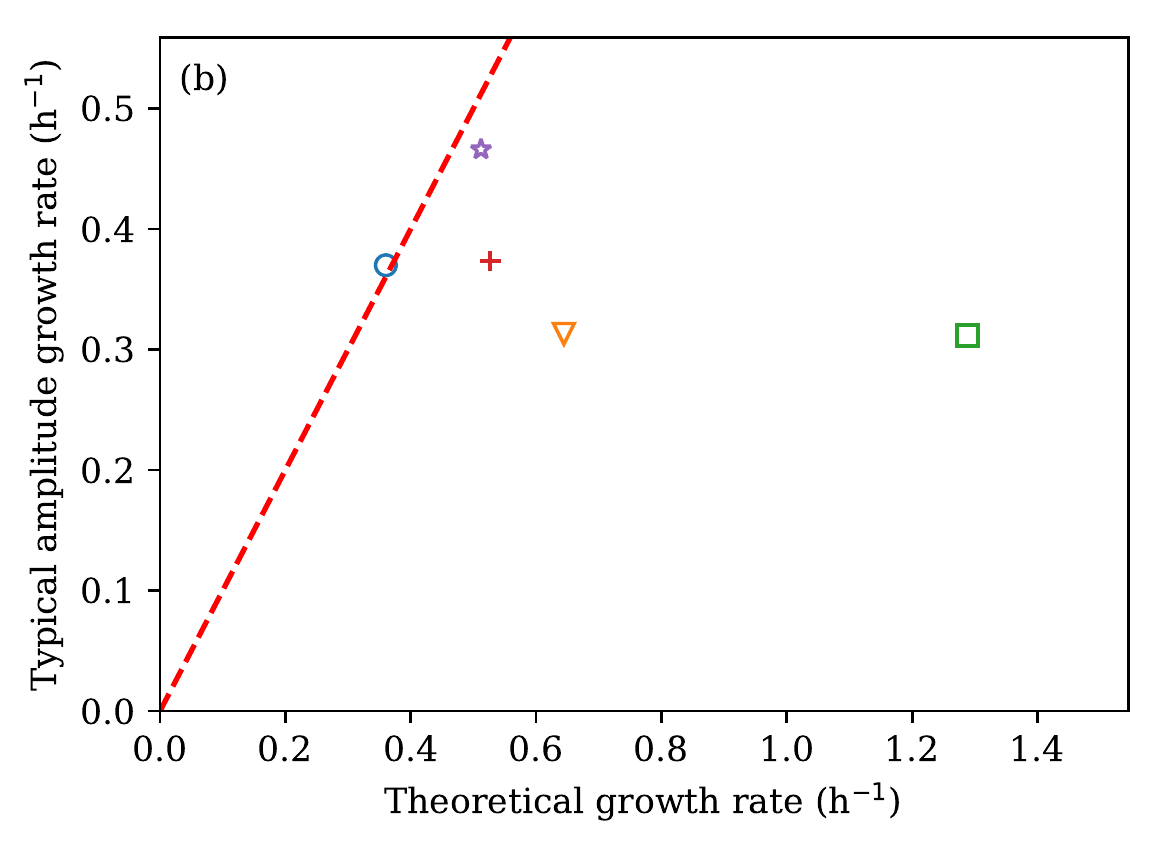}  
		  \label{subfig:WGR_U}
		\caption{{\bf (a)} Evolution of the typical amplitude (blue points) of the dissolution flutes during a typical experiment (run 3, $\circ$ discharge 2.8\,L\,min$^{-1}$, slope 39$^{\circ}$). The red curve is an exponential fit of the solid points in the left of the graph, corresponding to the first 4 hours of the experiment. The amplitude can be also satisfyingly fitted by a line (not shown) of slope $\mathrm{sl}=12$ $\mu$m h$^{-1}$. For the others measurement, we have $\mathrm{sl}=12$ $\mu$m h$^{-1}$ for run 1 $\square$, $\mathrm{sl}=18$ $\mu$m h$^{-1}$ for run 2 $\triangledown$, $\mathrm{sl}=14$ $\mu$m h$^{-1}$ for run 4 $+$ and $\mathrm{sl}=25$ $\mu$m h$^{-1}$ for run 5 $\star$. {\bf (b)} Depth growth rate of the flutes measured for the 5 experiments with respect to the theoretical growth rate as predicted in equation~(\ref{eq:growth_rate_prediction2}). The red dashed line corresponds to equality. {Note that the growth rate measurement depends on the interval on which the fit is performed, and should be considered as indicative.}}
	\label{fig:depth_growth_rate}
	\end{figure}  

Figure~\ref{fig:depth_growth_rate} (a) shows the evolution of the typical amplitude of the dissolution flutes during a typical experiment. The pattern amplitude grows nearly linearly with time, for $t \lesssim 5$ h. If the system obeys the prediction of the linear stability analysis, the amplitude should grow exponentially, at least at short times. To be self-consistent with this assumption, we select the data satisfying $t < 5$\,h (plain blue points), and fit this selected data with an exponential function (red line). We thus obtain the typical amplitude growth rate of the experiment, which we can compare to the theoretical growth rate $\sigma$ predicted by equation~(\ref{eq:growth_rate_prediction2}) with $\rho_{\mathrm{rock}}=\rho_{\mathrm{plaster}}=1200$ kg.m$^{-3}$ measured for our plaster blocks, $c_{sat}= 2.58 $ kg.m$^{-3}$ and $D=1.0\,10^{-9}$\,m$^2$\,s$^{-1}$~\cite{Colombani2007}. Fig.~\ref{fig:depth_growth_rate}~(b) shows this comparison for all five experiments made on plaster. We measure typical amplitude growth rates of order about 0.3-0.5~h$^{-1}$, which is close to the values predicted by the crude analysis presented here ( typically 0.3-1.3~h$^{-1}$). Yet, the analysis predicts that the growth rate should increase with the parameter $\sqrt{U_0}/h_0$, whereas our measurements rather tend to show that the growth rate is nearly constant and independent of this parameter, at least in the range of our experimental parameters.
 
In fact, the Figure~\ref{fig:depth_growth_rate} (a) shows a pattern amplitude evolution more compatible with a linear growth than with an exponential growth. As $\mathrm{exp} ( \gamma t) = 1+ \gamma \,t+ \frac{1}{2}\,(\gamma\, t)^2+ \ldots \,$, the linear behavior is valid only for $t \ll 2/\gamma$, which gives about $5$\,h with $\gamma \approx 0.4$~h$^{-1}$. Thus, the linear growth shown here at longer times cannot be attributed to the short-time behavior of an exponential growth. Usually in the study of hydrodynamic instabilities, additional nonlinear terms saturate the growth, but do not lead to a linear behavior to our knowledge. According to our experimental data, the dissolution pattern emergence is likely not explained by a classic linear instability mechanism. 
  
    \section{Response of the bed to transverse perturbations of the flow}
  \label{response}


A second kind of mechanism to explain emergence of a dissolution pattern consists in breaking the assumption of homogeneous flow along the spanwise axis at $t=0$. A flow heterogeneity can induce a differential erosion rate, and thus a dissolution pattern. We discuss in the discussion of the main document the possible origin of hydrodynamic streamwise structures like the turbulent streaks or the G\"ortler vortices. We assume the presence of a periodic modulation of the flow in space, $U=U_0\,[1+\beta\,\cos (k_\beta\,y)]$. By injecting this expression in the erosion rate previously derived (Eq.~(\ref{eq:erosion_rate})), we obtain an erosion rate depending on $y$. Supposing $\beta \ll 1$:
\begin{equation}
{\partial_t \eta}=-\dfrac{c_{sat}\,\sqrt{D\,U_0\,(1+\beta\,\cos (k_\beta\,y))}}{\rho_{rock}\,\sqrt{\pi\,x}}\approx V_{er,0}(x)\,\left(1+\dfrac{\beta}{2}\,\cos (k_\beta\,y)\right)
\end{equation}
The average erosion velocity and the growth velocity of the pattern read:
\begin{equation}
\left\lvert {\partial_t \eta_0}\right\rvert = V_{er,0}(x) \quad \mathrm{and}\quad \left\lvert {\partial_t \eta_1}\right\rvert = V_{er,0}(x)\,\dfrac{\beta}{2}\,\cos (k_\beta\,y)
\end{equation}
After time-integration and taking the standard deviation, one obtains the pattern amplitude $\sigma_\eta$, which evolves linearly with time:
\begin{equation}
\sigma_{\eta\,1}=\sigma_\eta=|V_{er,0}(x)|\,\dfrac{\sqrt{2}\,\beta}{4}\,t
\end{equation}

This simple models predicts that the pattern arises as a passive response to the flow heterogeneity. The pattern wavelength is directly imposed by the wavelength of the flow perturbation. The pattern amplitude increases linearly with time as observed in Fig.~\ref{fig:depth_growth_rate} (a) and is consequently proportional to the average eroded height. This model predicts a constant value of the ratio $\sigma_\eta/|\langle \eta\rangle|$ plotted in Fig.~{3} (d) in the main document. According to our measurements $\sqrt{2}\,\beta/4 \approx 0.1$, \textit{i.e.} $\beta \approx 0.3$. The pattern linear growth predicted by this model is characteristic of a passive response of the dissolving surface to the flow variations. We have also proposed a similar explanation of pattern emergence in our study of dissolution induced by solutal convection~\cite{Cohen2016,Philippi2019,Cohen2020}, but without reporting a linear growth of the pattern for now.

However this model is only justified at short times, because it neglects the feedback interaction of the pattern on the flow. Allen~\cite{Allen1970} proposes that the grooves channelize the  coherent structures. Due to the interaction with the topography, the $y$ position of the flow structures (like turbulent streaks) would be locked in time, which would explain the coherent growth of the pattern.  Moreover, the increase  of the typical width of the grooves is not captured by this simple model. Yet we note that, if $k_\beta$ is time-dependent whereas $\beta$ remains constant, the linear growth is conserved. The model is thus not incompatible with the pattern coarsening, but does not explain it. Qualitatively, a larger rill may focus a higher flow rate leading to a higher growth rate. If the grooves keep their aspect ratio during their growth, the small channels would be absorbed by the large ones, by a merging process. A more elaborated description able to describe the pattern evolution may combine both mechanisms. Such a model may incorporate simultaneously the response of the bed to the coherent hydrodynamic structures and the hydrodynamic feedbacks to the bed evolution.

%